\documentclass[floatfix,12pt,english]{iopart}
\usepackage[T1]{fontenc}
\usepackage[latin1]{inputenc}
\usepackage{graphicx}
\makeatletter
\usepackage{epsfig}
\usepackage{multicol}
\usepackage{latexsym}
\usepackage{amsfonts}
\makeatother
\usepackage{babel}
\def\be{\begin{equation}}
\def\ee{\end{equation}}
\newcommand{\beqa}{\begin{eqnarray}}
\newcommand{\eeqa}{\end{eqnarray}}

\newcommand{\omo}{\Omega_M^{(0)}}

\newcommand{\dl}{\delta}

\begin{document}

\title{Structure formation in the presence of 
dark energy perturbations}

\author{L. R. Abramo$^1$, R. C. Batista$^1$, L. Liberato$^2$
and R. Rosenfeld$^2$}

\address{$^1$ Instituto de F\'{\i}sica, Universidade de S\~ao Paulo\\
CP 66318, 05315-970, S\~ao Paulo, Brazil}
\address{$^2$ Instituto de F\'{\i}sica Te\'orica, Universidade Estadual Paulista\\
R. Pamplona 145, 01405-900, S\~ao Paulo, Brazil}

\eads{\mailto{abramo@fma.if.usp.br},
\mailto{rbatista@fma.if.usp.br},
\mailto{liberato@ift.unesp.br},
\mailto{rosenfel@ift.unesp.br}}

\begin{center}
\today
\end{center}

\begin{abstract}
We study non-linear structure formation in the presence of dark energy.
The influence of dark energy on the growth of large-scale
cosmological structures is exerted both through its background
effect on the expansion rate, and through its perturbations as well.
In order to compute the rate of formation of massive objects we
employ the Spherical Collapse formalism, which we generalize to
include fluids with pressure. We show that the resulting non-linear
evolution equations are identical to the ones obtained in the
Pseudo-Newtonian approach to cosmological perturbations, in the
regime where an equation of state serves to describe both the
background pressure relative to density, and the pressure 
perturbations relative to the density perturbations as well. 
We then consider a wide range of constant and time-dependent equations of
state (including phantom models) parametrized in a standard way, and
study their impact on the non-linear growth of structure. 
The main effect is the formation of dark energy structure associated
with the dark matter halo: non-phantom equations of state induce the formation of a 
dark energy halo, damping the growth of
structures; phantom models, on the other hand, generate dark energy voids,
enhancing structure growth.
Finally, we employ the Press-Schechter formalism to compute 
how dark energy affects the number of massive
objects as a function of redshift (number counts.)

\end{abstract}

\noindent{\it Keywords\/}: Cosmology: theory ---
Cosmology: large-scale structure of the Universe

\maketitle

\section{Introduction}

Observations of high-redshift SNIa imply that the expansion of the universe
has been accelerating in the past few billions of years
\cite{Knop:2003iy,Riess:2004nr,Astier:2005qq,Wood-Vasey:2007jb}.
This is corroborated by at least three broadly independent observations:
the angular spectrum of the cosmic microwave background temperature fluctuations
\cite{Spergel:2003cb,Spergel:2006hy},
the galaxy-galaxy correlation function, which traces the spatial distribution of
large-scale structure \cite{Tegmark:2003uf,Cole:2005sx}, and
the baryon acoustic oscillations \cite{BAO}.
Presently, the combined datasets favour a flat universe
with $\Omega_{\rm m} \simeq 0.27$, where the remaining $73\%$ of the energy
budget is taken up by dark energy.

These observations suggest that the dominant contribution to the present
energy density of the Universe can be described by a dark ({\it i.e.},
weakly or non-interacting) fluid with equation of
state (EoS) $w_{\rm de}=p_{\rm de}/\rho_{\rm de}<-1/3$. A particular case of
such a substance would
be the cosmological constant, $\Lambda$, for which $w_\Lambda=-1$.
Many other models with $w_{\rm de} \neq -1$ have been proposed,
usually in the framework of a scalar field (``quintessence") or some other
form of cosmic fluid with negative pressure -- see, e.g., \cite{Peebles:2002gy}
for a comprehensive review.

A more direct approach to the phenomenology of dark energy has been
recently adopted, in which the equation of state $w_{\rm de}$ is
expressed in terms of a certain parametrization with respect to
its time dependence
\cite{Linder:2002et,Sahni:2002fz,Alam:2003sc,Lazkoz:2005sp}.
Although determining the equation of state as a function of redshift would probably
not help to reveal the nature of dark energy, it could go a long way towards
discriminating among existing models.
Hence, one of the most important tasks ahead for observational cosmology is
to gather sufficient data to successfully and inequivocally distinguish
between this landscape of possibilities. As for theorists, the challenge is to
determine in which additional ways dark energy may manifest itself in nature,
apart from the acceleration of the overall expansion rate of the Universe.

One of the ways in which dark energy changes the evolution of our local
Universe is through its influence over the rates of formation and growth of
collapsed structures (halos). Since all galaxies and quasars, as
well as supernovae and putative sources of gamma-ray bursts, lie
inside collapsed structures of some type or another, their distribution
in size, space and in time will reflect
to some extent the influence of dark energy.

There are basically three mechanisms through which dark energy affects
large-scale structure. First,
the collapse of an overdense region due to gravitational instability
is slowed down by the Hubble expansion drag, so any additional
component which increases the expansion rate for the same value of the
energy density will dampen the formation of collapsed structures.
Second, as the accelerated expansion picks up speed,
the large-scale gravitational potentials grow slower, then start to decay.
This means that, as dark energy becomes the dominating dynamical component
of the Universe, some large-scale overdensities will grow slower,
and the process of gravitational collapse will even reverse itself
at scales comparable to the Hubble horizon.
And third, if dark energy is not the cosmological constant then it
must fluctuate both in time and in space. Hence, dark energy not
only feels the gravitational pull of a matter halo, but it tends
to form halos itself, thus influencing back those matter halos in a 
non-linear manner.
Notice that
the first two mechanisms affect collapsed structures only indirectly,
through changes in the Hubble expansion rate, while the third
mechanism depends on the clustering properties of dark energy.
Since different models of dark energy can easily produce the same
homogeneous expansion rate, but they hardly ever produce the same
perturbations, the largest potential to probe the nature of dark
energy possibly comes from such perturbative mechanisms.

In a previous paper, two of the present authors have studied the 
influence at the background level (no dark energy fluctuations)
of different parametrizations of the dark energy equation of state (EoS)
in the evolution of dark matter perturbations and in the final number
counts of dark matter halos \cite{Liberato:2006un}.
Our main purpose in this paper is to extend this analysis
by studying non-linear structure formation
including the possibility of dark energy fluctuations.

Related approaches were recently developed by 
Nunes \& Mota \cite{Nunes:2004wn}, 
Manera \& Mota \cite{Manera:2005ct},
Nunes, Silva \& Aghanim \cite{Nunes:2005fn}
and Dutta \& Maor \cite{Dutta:2006pn},
but those works considered scalar field dark energy.
Here we focus instead on dark energy as described by some
parametrization for the EoS as a function of redshift. 
This is more general than the scalar field approach, since the EoS
is directly related to the physical 
observables most widely used to measure cosmic acceleration.
Moreover, in contrast to \cite{Nunes:2004wn,Manera:2005ct},
we were able to investigate the non-linear regime of
both dark matter and dark energy clustering consistently, and we have found that 
it has an important effect on the formation of massive objects
($M>10^{13} M_{\odot}$).
Although the description of dark energy entirely in terms of its EoS 
may be too restrictive, the present work proves that the impact of dark energy 
perturbations on the formation of collapsed objects is both substantial and 
observable.

This paper is organized as follows. In Sec.~2 we extend the
traditional Spherical Collapse (SC) model,
originally used only to describe gravitational
collapse in the absence of pressure, to incorporate
the possibility of coupled perturbations in 2 fluids,
namely pressureless
dark matter and negative pressure dark energy.
We verify this generalization showing
under what conditions the SC model is equivalent to
a Pseudo-Newtonian (PN) perturbation theory approach.
In Sec.~3 we study
the linear evolution of the generalized SC equations, and the effects
of dark energy on the formation and on the initial stages of the
evolution of matter halos.
In Sec.~4 we analyze the fully non-linear system of SC equations,
and how the formation of strongly non-linear (collapsed) matter
halos both affects and is affected by dark energy halos.
In Sec.~5 we show how our results can be included in a
Press-Schechter formalism in order to derive the consequences
to dark matter halos number counts. We conclude in Sec.~6.

\section{The spherical collapse model and its generalizations}

The simplest (semi-) analytical tool to study non-linear structure formation is the
SC model \cite{Gunn:1972sv}.
It has been shown that the SC equations
can be actually derived from General Relativity, as long as shear does
not play a significant role \cite{Gaztanaga:2000vw}.

Most studies about the impact of dark energy on structure formation were
performed under the assumption that dark energy is uniformly distributed.
In this case, where dark energy affects only background quantities,
the SC model can be easily modified to incorporate dark energy effects.
For instance,
the abundance of rich clusters of galaxies estimated within the SC model
was used to constrain the cosmological
model and the properties of dark energy fluid in the context of the simplest
case of a cosmological constant \cite{Lahav:1991wc,Viana:1995yv}, in 
the case of a constant $w_{\rm de}\neq-1$
\cite{Wang:1998gt,Lokas:2001nw,Basilakos:2003bi,Percival:2005vm,Horellou:2005qc},
as well as the case of dynamical dark energy models with some parametrizations
of $w_{\rm de}(t)$ \cite{Liberato:2006un}.

However, the standard SC framework was originally designed to 
describe perturbations
in pressureless matter, while we are interested in the effects
of perturbations in an extra component whose pressure is 
very large and negative.
If we want to study a gravitationally coupled system of matter (we
do not distinguish between dark
matter and baryons here) and dark energy, the SC model must be expanded
beyond the realm of the Einstein-de Sitter model.

Consequences of dark energy fluctuations in the
studies of structure formation are more naturally incorporated
by introducing a scalar field with a suitable potential to model
the dark energy component,
such as the quintessence field.
In this approach, the authors of Refs.
\cite{Nunes:2004wn,Manera:2005ct,Maor:2005hq,Mota:2004pa}
proposed an extension of the SC equations that take into 
account fluctuations in the
dark energy field for minimally and non-minimally coupled 
quintessence field.

It is often more convenient,
and completely equivalent at the background level, 
to introduce a time-dependent
parametrization for the dark energy EoS, $w_{\rm de}(z)$.
Since it is possible to reconstruct the scalar field potential
from a general parametrization of dark energy, or directly
from the EoS $w_{\rm de}(z)$ 
\cite{Guo:2005at,Rosenfeld:2007ri},
the two approaches are in fact closely related.
Our goal in this section is to generalize the SC 
model with a fluid description
of dark energy in order to include the possibility 
of dark energy fluctuations.

We will check this generalized SC model with the results from
the Pseudo-Newtonian (PN) approach to cosmology \cite{Lima:1996at}
for perfect fluids with pressure.
The advantages of the PN framework are that it
is both simpler than full-blown non-linear
General Relativity (GR), and more intuitive.
Crucially, it is in good agreement with
GR in the linear regime  \cite{Hwang:1999yv,Reis:2003fs}.
As we will see, the PN approach is also particularly useful if 
we want to keep contact with
the description of dark energy in terms of a parametrization for its
EoS, and it can be easily generalized
to a multi-fluid system.

The only remaining question is whether the two approaches agree
with each other.
Next we verify under which conditions
the SC model is equivalent to the PN approach.

\subsection{Spherical Collapse}

The continuity equation for a single perfect fluid $j$ with background 
density $\rho_j$ and pressure $p_j=w_j  \rho_j$ is given by:
\begin{equation}
\dot{\rho_j}+3H\rho_j\left(1+w_j\right)=0\;,\label{cont}
\end{equation}
where $H=\dot{a}/a$ is the Hubble parameter.
Consider now
a spherically symmetric region of radius $r$ and with
a homogeneous density $\rho_{c_{j}}$ (a top-hat distribution).
Suppose that, at time $t$, $\rho_{c_j}(t) = \rho_j(t) + \delta\rho_{j}$.
If $\delta\rho_{j} >0$ this spherical region will eventually
collapse from its own gravitational pull,
otherwise it will expand faster than the average Hubble flow,
generating what is known as a void.
The evolution of such simplified spherical regions
can be described in close analogy with the continuity Eq.~(\ref{cont}),
but now with $p_{c_j}=w_{c_j}\rho_{c_j}$:
\begin{equation}
\dot{\rho}_{c_j}+3h\rho_{c_j}\left(1+w_{c_j}\right)=0 \; ,
\label{cont-coll}
\end{equation}
where $h=\dot{r}/r$ denotes the local expansion rate inside
the spherical region. Note that, in principle, we could have different
equations of state inside and outside the spherical region,
$w_{c_j}\neq w_j$.
In fact, the difference between the local and the
background equations of state $\delta w_j \equiv w_{c_j}- w_j$ 
can be related to the
fluid's effective speed of sound, $c_{{\rm eff} \, j}^{2}=\delta
p_j/\delta\rho_j$, 
through:
\begin{equation}
\delta w_j = \frac{\delta \rho_j}{\rho_j + \delta \rho_j } (c_{{\rm eff} \, j}^{2} - w_j).
\end{equation}
Usually, $c_{{\rm eff}}^2$ is regarded as a free parameter
-- although, rigorously, in perturbation theory the only other free parameter 
is the true sound speed of inhomogeneities, $c_X^2$ \cite{Garriga:1999vw}.
Here the sound speed $c_{{\rm eff}}^2$ is defined as the ratio 
between two independent perturbative degrees of freedom, 
so not only it is gauge dependent, but it may
depend also on the initial conditions for those 
perturbations. Therefore, $c_{{\rm eff}}^2$ stands as a proxy for
the pressure perturbations. 

For simplicity, and in order to make contact with the PN equations,
we will consider the case where the EoS 
is the same inside the collapsing sphere and in the background, 
so we take $\delta w_j =0$ and thus $c_{{\rm eff} \, j}^{2} = w_j$.
This situation can be readily obtained in cases such as a slow-rolling
scalar field.

It should be noted that, in principle, there are instabilities in 
the growth of inhomogeneous perturbations whenever the sound speed becomes negative.
However, within the spherical collapse model with a top-hat profile and
the assumption of a space-independent $c_{{\rm eff}}^2$,
there are no pressure or density gradients, so no such problem of
instabilities arises.

By the same token as the first Friedmann equation, 
consider now the second Friedmann equation 
applied to the spherical region:
\begin{equation}
\label{Fried}
\frac{\ddot{r}}{r}=-\frac{4\pi G}{3}\left(\rho_{c}+3p_{c}\right) \; .
\end{equation}
Notice that the density and pressure that appear in Eq.
(\ref{Fried}) are the sum of densities and pressures
of {\it all} contributing fluids, while the continuity
Eqs.~(\ref{cont})-(\ref{cont-coll})
are valid for each individual fluid (in the absence, of course,
of direct couplings between those fluids.)

It is useful to define the density contrast of a single
fluid species $j$ by the relation:
\begin{equation}
\delta_{j}+1=\frac{\rho_{c_{j}}}{\rho_{j}}\;.
\end{equation}
Differentiating this with respect to time we obtain:
\begin{equation}
\dot{\delta_{j}}=3(1+\delta_{j})
(H - h) (1+w_{j}) \; ,
\label{dponto}
\end{equation}
where we assumed $w_{c_{j}}=w_{j}$. Differentiating again with respect to
time and employing the equations for the background and for the spherical
region, we can derive the following non-linear evolution equation for $\delta_{j}$:

\begin{eqnarray}
\ddot{\delta}_{j}+\left(2H-\frac{\dot{w}_{j}}{1+w_{j}}\right)\dot{\delta}_{j}-
4 \pi G \left(1+w_{j}\right)\left(1+\delta_{j}\right)
\sum_{k}\rho_{k}\delta_{k}\left(1+3w_{k}\right)= \nonumber \\
\left[\frac{4+3w_{j}}{3(1+w_{j})}\right]
\frac{\dot{\delta}_{j}^{2}}{1+\delta_{j}} \; .
\label{SC_eq}
\end{eqnarray}

Notice that we admit the possibility of a time-dependent EoS.
For a system of $n$ fluids, we must consider $n$ equations such as
(\ref{SC_eq}), all coupled gravitationally through the term proportional to
Newton's constant. Although they are not derived rigorously from
General Relativity, we will see next that these equations
find support in the PN approximation to gravitational interactions.

\subsection{Pseudo-Newtonian Cosmology}

Consider now the PN cosmological model, described by the
equations \cite{Lima:1996at}:
\begin{equation}
\frac{\partial\rho _j}{\partial t}+\vec{\nabla}_{r}\cdot\left(\vec{u}_j\rho_j\right)+p_j\vec{\nabla}_{r}\cdot\vec{u}_j=0\;,\label{cont-pnc}
\end{equation}
\begin{equation}
\frac{\partial\vec{u}_j}{\partial t}+\left(\vec{u}_j\cdot\vec{\nabla_{r}}\right)\vec{u}_j=-\vec{\nabla}_{r}\Phi-\frac{\vec{\nabla}_{r}p_j}{\rho _j+p_j}\;,\label{euler-pnc}
\end{equation}
\begin{equation}
\nabla_{r}^{2}\Phi=4\pi G\sum_k\left(\rho_k+3p_k\right)\;,\label{poison-pnc}
\end{equation}
where $\rho_j$, $p_j$, $\vec{u}_j$ and $\Phi$ denote, respectively, the density,
pressure, velocity and the
Newtonian gravitational potential of the cosmic fluid. 
These equations
are, respectively, generalizations of the continuity equation, of 
Euler's equation (both valid for each fluid species $j$), and of Poisson's 
equation (which is valid for the sum of all fluids.)

Cosmological perturbations are introduced by admitting
inhomogeneous deviations away from the background quantities:
\begin{eqnarray}
\rho _j & = & \rho_{0_j}(t)+\delta\rho_j(\vec{x},t) \; ,
\label{rho-pert}\\
p_j & = & p_{0_j}(t)+\delta p_j (\vec{x},t) \; , \\
\vec{u} _j & = & \vec{u}_{0_j}(t)+\vec{v}_j (\vec{x},t) \; , \\
\Phi & = & \Phi_{0}(t)+\phi (\vec{x},t)\; .
\end{eqnarray}

Changing to comoving coordinates, $\vec{x}=\vec{r}/a$,
(henceforth $\vec{\nabla}$ refers to gradient with respect to
comoving coordinates $\vec{x}$) and
using $\delta_j=\delta\rho_j/\rho_{0_j}$, we find the following equations
for the perturbed quantities:
\begin{equation}
\dot{\delta}_j+3H\left(c_{{\rm eff} \, j}^{2}-w_j\right)\delta_j
=-\left[1+w_j+\left(1+c_{{\rm eff} \, j}^{2}\right)\delta_j\right]
\frac{\vec{\nabla}\cdot\vec{v}_j}{a}
-\frac{\vec{v}_j\cdot\vec{\nabla}\delta_j}{a}
\label{cont-pert2}
\end{equation}
\begin{equation}
\dot{\vec{v}_j}+H\vec{v}_j+\frac{\vec{v}_j\cdot\vec{\nabla}}{a}\vec{v}_j
=-\frac{\vec{\nabla}\phi}{a}
-\frac{c_{{\rm eff} \, j}^{2}\vec{\nabla}\delta}
{a\left[1+w_j+(1+c_{{\rm eff} \, j}^{2}) \delta_j\right]}\;,
\label{euler-pert2}
\end{equation}
\begin{equation}
\frac{\nabla^{2}\phi}{a^{2}}
=4\pi G\sum_k\rho_{0_k}\delta_k\left(1+3c_{{\rm eff} \, k}^{2}\right)\;,
\label{poisson-pert2}
\end{equation}
where $c^2_{{\rm eff} \, j} \equiv \delta p_j/\delta\rho_j$ is the 
effective sound speed of each fluid.
In order to obtain these equations we have
assumed that $w_j$ and $c_{{\rm eff} \, j}^{2}$ are functions of time only.

Notice that Eqs.~(\ref{cont-pert2})-(\ref{poisson-pert2})
are valid even if $\delta_j$ is not small, so we can use them to
follow the evolution of a collapsing region well into the non-linear
regime. In fact, the PN equations of
motion become a better approximation as the size of the system
shrinks due to gravitational collapse. This is easy to see by
noticing that in most collapsed regions of the Universe the
density contrast $\delta_j$ may be extremely large, but the
gravitational potentials are small, $\phi \ll 1$,
and the local relative (peculiar) velocities are almost never relativistic.
Hence, the PN equations
may be a poor approximation at the moment of turnaround (when
a spherical region breaks away from the Hubble flow) for
the scales comparable to the Hubble horizon at the time of
turnaround, but for all other scales and epochs it is a good approximation
that becomes progressively better as the system collapses.

In order to simplify the PN equations, it is useful to define:
\begin{equation}
\theta_j
\equiv
\vec{\nabla}\cdot\vec{v}_j
\; ,
\end{equation}
\begin{equation}
C_j\equiv a^{-1}\vec{\nabla}\cdot\left[\left(\vec{v}_j\cdot\vec{\nabla}\right)\vec{v}_j\right]\;,
\end{equation}
and
\begin{equation}
f_j\equiv\vec{\nabla}\cdot\left[\frac{\vec{\nabla}\phi}{a}+\frac{c_{{\rm eff} \, j}^{2}\vec{\nabla}\delta_j}{a\left(1+w_j+\delta_j+c_{{\rm eff} \, j}^{2}\delta_j\right)}\right]\;,
\end{equation}
so that by taking the divergence of (\ref{euler-pert2}) we obtain:
\begin{equation}
\dot{\theta}_j+H\theta_j+C_j=-f_j\;.\label{euler_sh}
\end{equation}
We also define:
\begin{equation}
A_j\equiv3H\left(c_{{\rm eff} \, j}^{2}-w_j\right)\delta_j \; ,
\end{equation}
\begin{equation}
B_j\equiv1+w_j+\left(1+c_{{\rm eff} \, j}^{2}\right)\delta_j\;,
\end{equation}
and by neglecting the term $\vec{v_j}\cdot\vec{\nabla}\delta_j$, which is
of order of $v^{2}_j/c^{2}$, we can cast Eq.~(\ref{cont-pert2}) in
the form:
\begin{equation}
\dot{\delta}_j+A_j+\frac{\theta_j}{a}B_j=0\;.\label{cont_sh}
\end{equation}
Taking the partial derivative of (\ref{cont_sh}) with respect to
time, using Eq.~(\ref{euler_sh}) to eliminate $\dot{\theta}_j$ and
Eq.~(\ref{cont_sh}) to eliminate $\theta_j$ we get:
\begin{equation}
\ddot{\delta}_j+\dot{A}_j+\left(A_j+\dot{\delta}_j\right)\left(2H-\frac{\dot{B}_j}{B_j}\right)-\frac{B_j}{a}\left(f_j+C_j\right)=0.
\label{newt_sc}
\end{equation}
In the appendix we explicitly show that, for a single
fluid with $c_{eff}^2 = w = const$, PN and GR at the linear level
differ only by a decaying mode.   

Now we try to make contact with the SC equations.
In order to reproduce Eq.~(\ref{SC_eq})
we must, first of all, assume that the velocity 
profile is consistent with the hypothesis of spherically 
symmetric collapse of a top-hat
inhomogeneity, i.e., $\vec{v}_j= \theta(t)/3 \, \vec{x}$.
Second, we also have to assume that 
$c_{{\rm eff} \, j}^{2}=w_j$. 
Notice that because the intrinsic non-adiabatic pressure
$\Gamma_j \sim \delta p_j - c^2_{s\;j} \delta \rho_j$,
where the adiabatic sound speed is $c^2_{s\;j} = \frac{ \dot{p}_j}{\dot{\rho}_j}$,
our choice implies some amount of intrinsic entropy perturbations
for the dark energy fluid.

With these choices we have: $A_j=0$, $B_j=\left(1+w_j\right)\left(1+\delta_j\right)$
and $C_j=\theta^{2}_j/3a$. Notice that, with this velocity
field, the LHS of equation (\ref{euler_sh}) is identical to that
of Raychaudhuri's equation when we assume that $\theta_j$ is a function
of time only:
\begin{equation}
\dot{\theta}_j+H\theta_j+\frac{\theta^{2}_j}{3a}=-f_j \; .
\end{equation}
This equation reduces to the one found in \cite{Gaztanaga:2000vw} if
we neglect the gradients of the density contrast in $f_j$.
Since we are considering the spherical collapse of a
top-hat distribution (which is homogeneous inside the
radius $r$), the terms including $\vec\nabla \delta_j$
which appear in $f_j$ vanish. Under these conditions,
Eq.~(\ref{newt_sc}) reduces to the equation for SC, Eq.~(\ref{SC_eq}).

It is interesting that in fact we were forced to assume 
both that $w_{c_j}=w_j$ in
the SC formalism, and that $c_{{\rm eff} \, j}^{2}=w_j$ in
the PN formalism, in order that the two frameworks would result in 
identical equations.
This is a further motivation for our choices of
$\delta w_j=0$ and $c_{{\rm eff} \, j}^{2}=w_j$: only in this
scenario we can trust that the physics of non-linear 
spherically symmetric collapse is well described
by our dynamical equations. In order to describe a more general 
situation probably neither approach is suited, and one would be 
forced to resort to full-blown General Relativity.
However it is possible that the numerical differences between
SC and PN for other choices of $\delta w_j$ and $c_{{\rm eff} \, j}^{2}$
are small.

\subsection{Equations for non-linear spherical collapse in the presence of
  dark energy}

We obtained non-linear differential equations that characterize 
the growth of spherically symmetric perturbations in fluids with 
arbitrary time-dependent equations of state. These equations are
coupled through the gravitational interactions. 
We saw that both the PN and SC approaches agree 
with General Relativity for a pressureless fluid; 
furthermore, we have shown that 
they agree with each other in the case where 
$c_{{\rm eff} \,  j}^{2}=w_{c_j}=w_j$ and if the density profile is a top-hat
($\vec\nabla \delta_j = 0$ .)

Particularizing to a model with only non-relativistic 
matter and dark energy,
in which the latter is characterized solely by
its EoS, the top-hat spherical
regions evolve according to a system of equations equivalent
to (\ref{SC_eq}):
\begin{eqnarray}
\ddot{\delta}_{\rm m}
+2H\dot{\delta}_{\rm m}
-\frac{4\dot{\delta}_{\rm m}^{2}}{3\left(1+\delta_{\rm m}\right)}=
\frac{3H^{2}}{2}\left(1+\delta_{\rm m}\right)\left[\Omega_{\rm m}\delta_{\rm m}
+\Omega_{\rm de}\delta_{\rm de}\left(1+3w_{\rm de}\right)\right]\;,
\label{system1} \\
\ddot{\delta}_{\rm de}
+\left(2H-\frac{\dot{w}_{\rm de}}{1+w_{\rm de}}\right)\dot{\delta}_{\rm de}
-\left[\frac{4+3w_{\rm de}}{3(1+w_{\rm de})}\right]
\frac{\dot{\delta}_{\rm de}^{2}}{1+\delta_{\rm de}}=
\nonumber \\
\frac{3H^{2}}{2}\left(1+w_{\rm de}\right)\left(1+\delta_{\rm de}\right)
\left[\Omega_{\rm m}\delta_{\rm m}+\Omega_{\rm de}\delta_{\rm de}
\left(1+3w_{\rm de}\right)\right] \; ,
\label{system2}
\end{eqnarray}
where $\delta_{\rm m}$ is the density contrast in matter and
$\delta_{\rm de}$ is the density contrast in the dark energy component.
These are the equations we will study in the following sections.

\section{Solutions in linear regime}

The linear regime of cosmological perturbations is valid for all
scales during the radiation era, and for most scales during the matter
era up until very recently. The initial stages of the process
of gravitational collapse are indeed very well described by the
linear regime for all but the smallest scales. Since this is
a simple system which can be studied almost entirely with
analytical tools, it is useful to try and extract some physics
from Eqs.~(\ref{system1})-(\ref{system2}) while they are
still in the linear regime. Neglecting the ${\cal{O}}(\delta^2)$ 
terms we obtain:
\begin{equation}
\ddot{\delta}_{\rm m}+2H\dot{\delta}_{\rm m}
=\frac{3H^{2}}{2}\left[\Omega_{\rm m}\delta_{\rm m}
+\Omega_{\rm de}\delta_{\rm de}\left(1+3w_{\rm de}\right)\right]\;,
\label{lin-mat}
\end{equation}
\begin{eqnarray}
\label{lin-de}
\ddot{\delta}_{\rm de}+\left(2H-\frac{\dot{w}_{\rm de}}
{1+w_{\rm de}}\right)\dot{\delta}_{\rm de}
\\ \nonumber
=
\frac{3H^{2}}{2}\left(1+w_{\rm de}\right)
\left[\Omega_{\rm m}\delta_{\rm m}+\Omega_{\rm de}\delta_{\rm de}
\left(1+3w_{\rm de}\right)\right]\;.
\end{eqnarray}

In principle, we can employ any given parametrization for dark
energy as a function of time or redshift, but in order to find closed
analytical formulas we initially take ${w}_{\rm de}=$constant.
We start by solving Eqs.~(\ref{lin-mat})-(\ref{lin-de}) well
inside the matter-dominated period ($z = 10^{3}$), when it
is a good approximation to assume that
$\Omega_{\rm de}\approx0$ and $\Omega_{\rm m}\approx1$.
Changing the time variable to the scale factor $a$, the equations
become:
\begin{eqnarray}
{\delta}_{\rm m}''+\frac{3}{2}\frac{{\delta}_{\rm m}'}{a}
-\frac{3}{2a^{2}}\delta_{\rm m} & = & 0
\;,\label{m-a-lin}\\
{\delta}_{\rm de}''+\frac{3}{2}\frac{{\delta}_{\rm de}'}{a}
-\frac{3}{2a^{2}}\left(1+w_{\rm de}\right)\delta_{\rm m} 
& = & 0\;,\label{de-a-lin}
\end{eqnarray}
where a prime denotes derivative with respect to $a$.

As is widely known, in this case the solution for the matter density
contrast is $\delta_{\rm m}\left(a\right)=C_{1} \, a+ C_{2} \, a^{-3/2}$,
where $C_1$ and $C_2$ are arbitrary constants.
Neglecting the decaying mode of the matter density contrast,
Eq.~(\ref{de-a-lin}) then has the solution:
\begin{equation}
\delta_{\rm de}= C_1 \left(1+w_{\rm de}\right) \, a + C_{3}
= \left(1+w_{\rm de}\right) \delta_{\rm m} + C_3 \;.
\label{de-sol}
\end{equation}

It is interesting to note that the adiabatic condition 
is $\delta_{\rm de}=\left(1+w_{\rm de}\right)\delta_{\rm m}$.
Hence, any value $C_{3}\neq0$ implies a
non-adiabatic initial condition -- i.e., in such a case 
the perturbations have an isocurvature component.
However, local non-adiabatic perturbations 
are unstable, since they correspond to pressure gradients 
between the internal and external parts of the
spherical region. These pressure gradients 
must eventually cancel out, so any
non-adiabatic component must decay during the evolution of 
the perturbations, leaving
only the usual adiabatic (curvature) fluctuations.

Notice that the condition of adiabaticity is different in the
usual dark energy models ($w_{\rm de} > -1$) compared to phantom models
($w_{\rm de} < -1$): for phantom models, adiabatic initial conditions
mean that any initial overdensity in matter is matched by an 
underdensity in dark energy, and vice-versa.
So, for example, take a phantom dark energy model and
a positive density perturbation in dark matter. If initially the
dark energy perturbation is also positive, then the pressure
gradients will cause the dark energy halo to decay, then to
turn it into a void, thus switching the sign of the dark energy 
perturbation -- see \cite{Maor:2005hq} for a similar
switching effect.

The effect of dark energy perturbations on
the evolution of dark matter perturbations is easily
understood from Eq.~(\ref{lin-mat}): dark energy perturbations
become a source for dark matter perturbation. Since 
$\left(1+3w_{\rm de}\right)<0$, a dark energy overdensity 
decreases dark matter clustering, which is intutitive
since a local concentration of dark energy would 
speed up the acceleration in that region.
The opposite holds for a region with a dark energy underdensity.
We show some examples in the next subsections. For the remainder
of the paper we adopt $\Omega_m = 0.25$,  $\Omega_{de} = 0.75$ 
and $h_0 = 0.72$.

\subsection{Constant $w_{\rm de}$: non-phantom models}

In the case of non-phantom dark energy we have $1+w_{\rm de}>0$,
and therefore Eq.~(\ref{de-sol}) implies that a region containing a
matter overdensity ($\delta_{\rm m} >0$) induces a dark energy overdensity 
($\delta_{\rm de}>0$) in that same region. Conversely,  a matter underdensity
region ($\delta_{\rm m} <0$) induces a dark energy underdensity 
($\delta_{\rm de}<0$). Hence, in the non-phantom case
a halo of dark matter induces a halo of
dark energy, and a void of dark matter induces a void of dark energy.
This behaviour is generic if we limit the scope of initial conditions to
adiabatic perturbations -- as predicted by inflation and confirmed by
WMAP \cite{Spergel:2006hy}.

\begin{figure}[htb]
\includegraphics[scale=0.31]{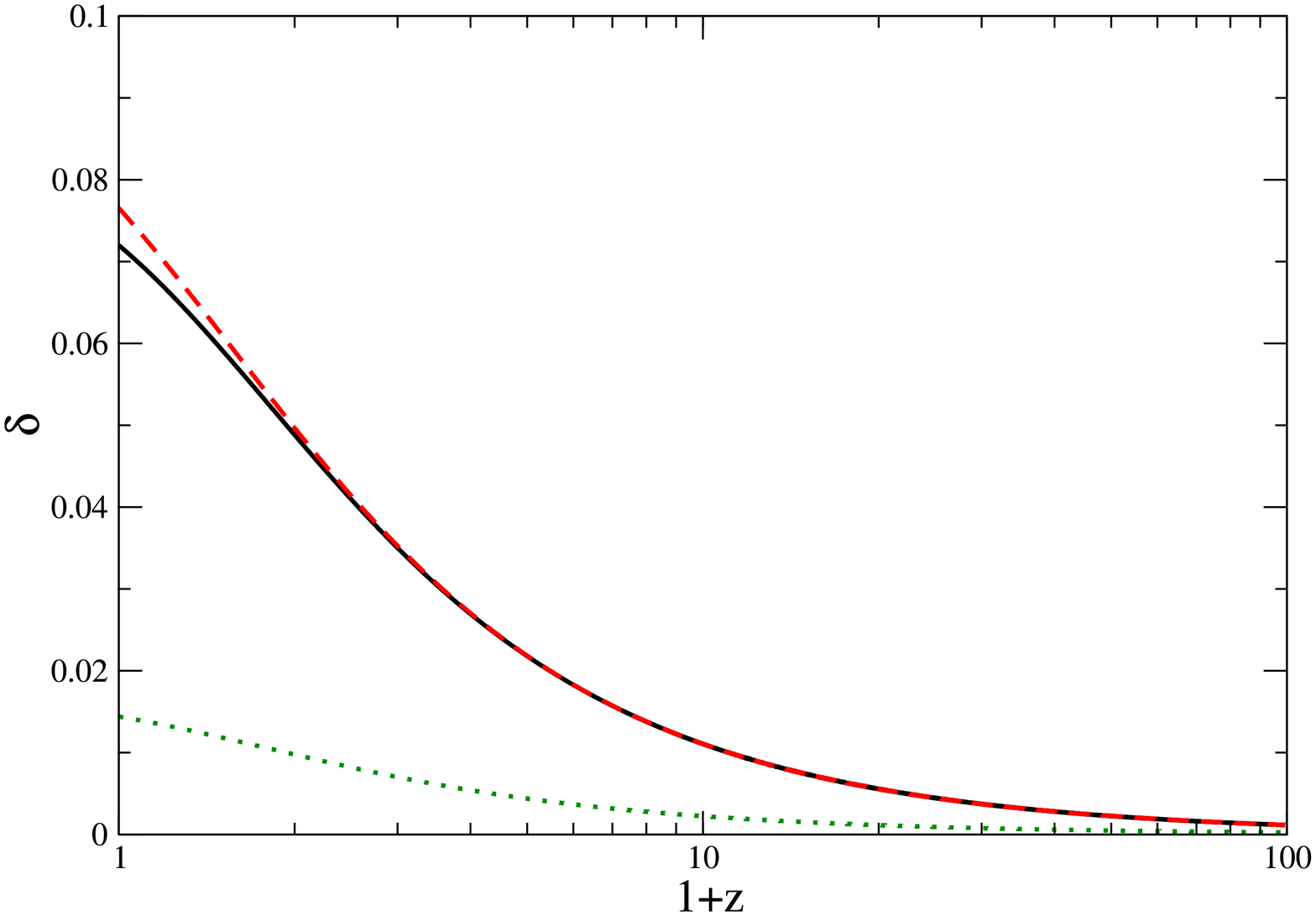}
\hspace{-1cm}
\includegraphics[scale=0.31]{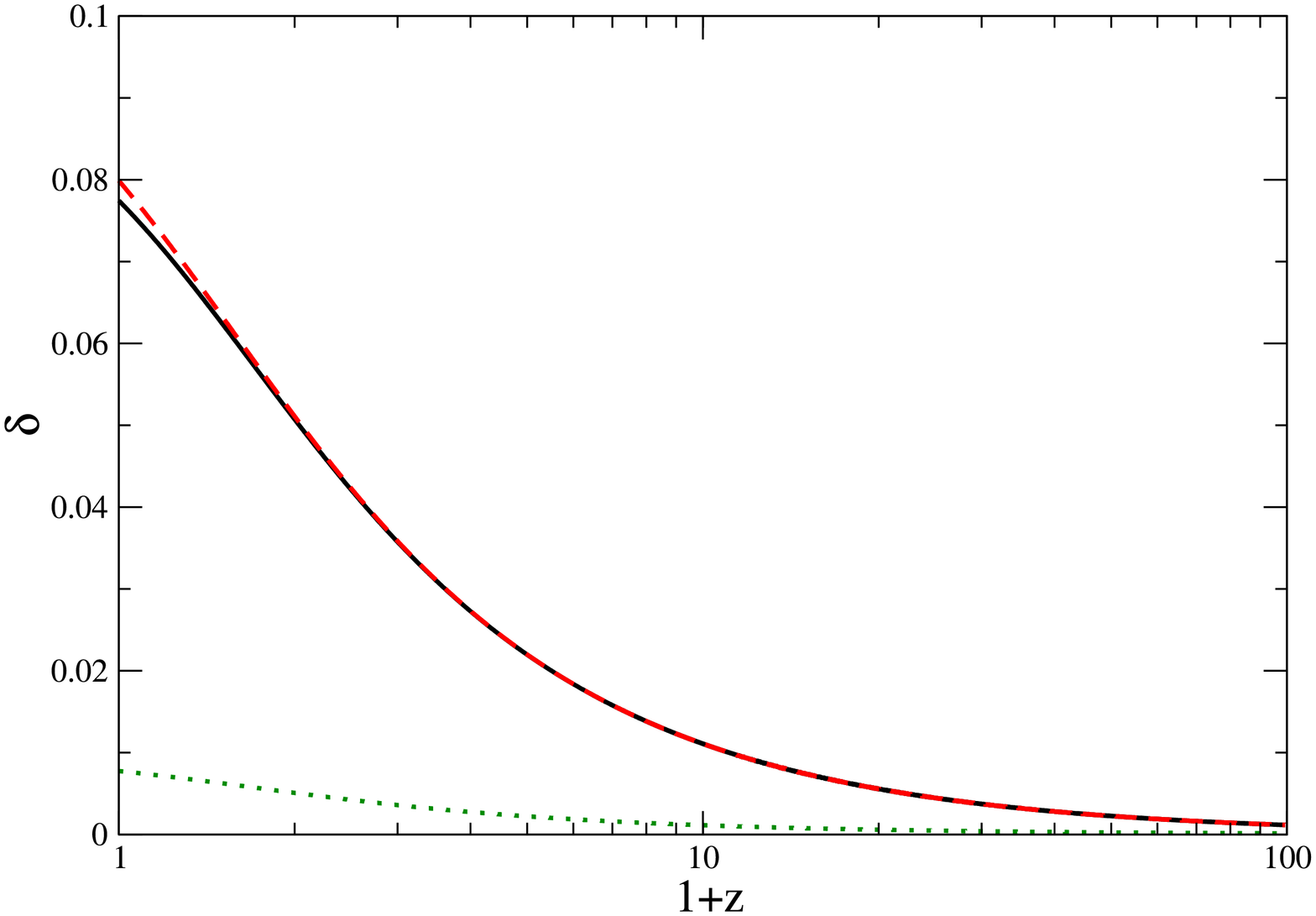}
\vspace{-1cm}
\caption{\sf Growth of matter perturbation with (solid lines)
and without (dashed lines) dark energy perturbations together with
growth of dark energy perturbations (dotted lines) for
$w_{\rm de} = -0.8$ (left panel) and $w_{\rm de}-0.9$ (right panel).
\label{non_phan_lin}}
\end{figure}

In order to study the impact of dark energy fluctuations 
on the growth of dark matter
perturbations we show in Fig.~\ref{non_phan_lin} the 
evolution of $\delta_{\rm m}$
with and without the inclusion of dark energy fluctuations, for 
two different values of
the EoS: $w_{\rm de} = -0.9$ and $-0.8$. We use adiabatic initial 
conditions at $z_i=1000$,
with $\delta_{\rm m}'(z_i) = -\delta_{\rm m}(z_i) /(1+z_i)$. 

The initial condition
on the derivative of the density perturbation comes from the 
assumption that dark energy
is negligible initially. As the figure shows, the inclusion of dark 
energy perturbations actually suppresses
the growth of dark matter perturbations in this case.
The differences in the linear regime are
roughly proportional to $1+w_{\rm de}$.
In this same figure we also show the growth of 
the dark energy perturbation, which is
much smaller than the matter fluctuation, 
as expected, but also tends to form a
dark energy halo.

\subsection{Constant $w_{\rm de}$: phantom models}

In the case of phantom dark energy we have $1+w_{\rm de}<0$,
and therefore Eq.~(\ref{de-sol}) implies that a matter overdensity
region ($\delta_{\rm m} >0$), which will later become a dark matter halo,
induces a dark energy density void ($\delta_{\rm de}<0$), and vice-versa.
Again, this behaviour is generic for purely adiabatic initial conditions.

\begin{figure}[htb]
\includegraphics[scale=0.31]{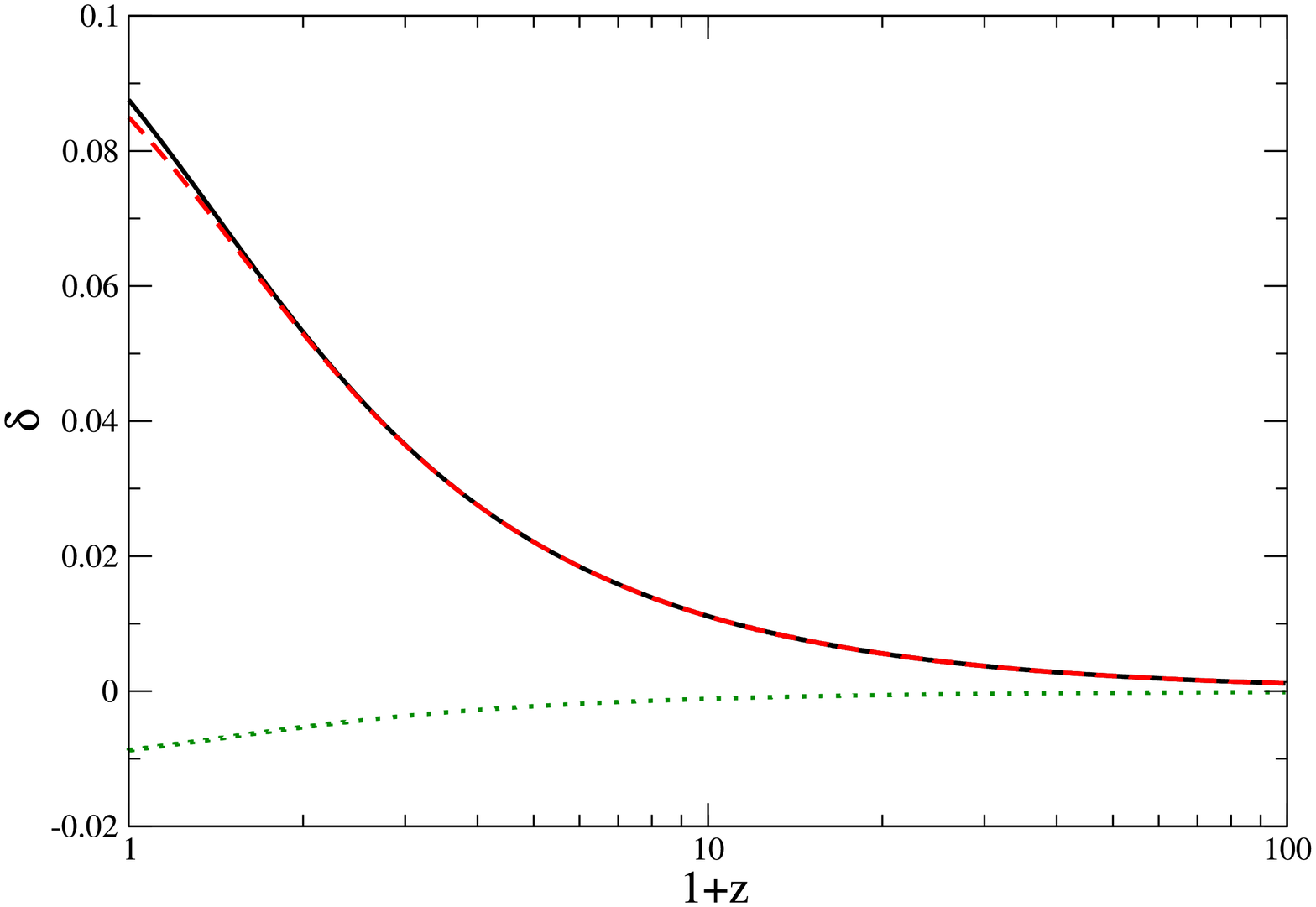}
\hspace{-1cm}
\includegraphics[scale=0.31]{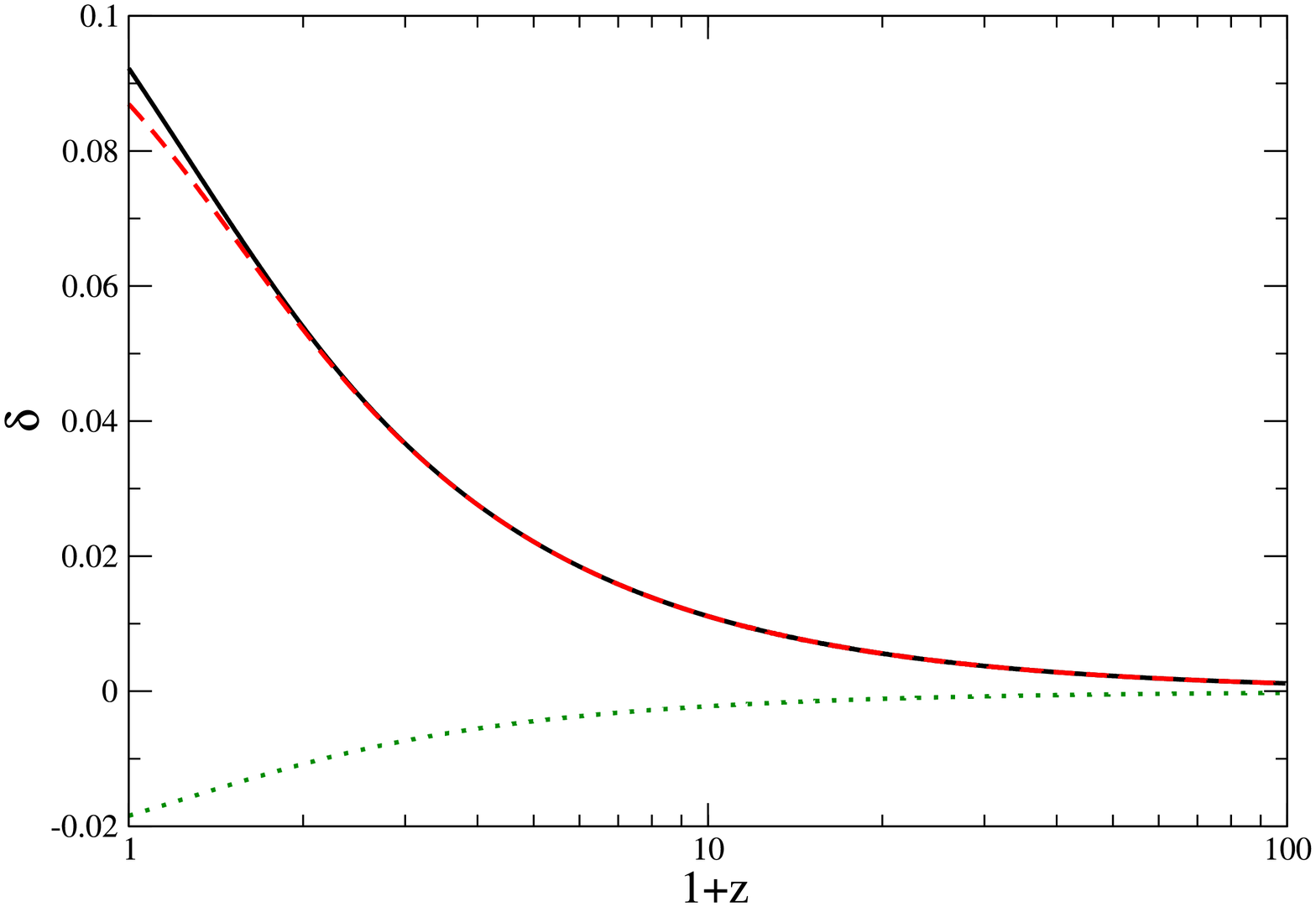}
\vspace{-1cm}
\caption{\sf Growth of matter perturbation with (solid lines)
and without (dashed lines) dark energy perturbations together with
growth of dark energy perturbations (dotted lines)  for
$w_{\rm de} = -1.1$ (left panel) and $w_{\rm de}=-1.2$ (right panel).
\label{phan_lin}}
\end{figure}

In Fig.~\ref{phan_lin} we show the effects of dark energy fluctuations
on the growth of dark matter perturbations for two different values of
the EoS, $w_{\rm de} = -1.1$ and $-1.2$.
As in the case of non-phantom dark energy, the differences are small
and increase with larger values of  $|1+w_{\rm de}|$.
However, contrary to the non-phantom case, fluctuations in phantom dark energy
enhance the growth of dark matter perturbations.
We also show the growth of the dark energy perturbation which, as expected,
tends to form a dark energy void.

\subsection{Varying $w_{\rm de}$}
\label{variable}

In the framework of single scalar field descriptions of 
dark energy it is impossible
for the EoS to cross the so-called phantom barrier at $w_{\rm de}=-1$
\cite{Vikman:2004dc}.
However, in our phenomenological approach we could in principle 
have a time-varying
parametrization of $w_{\rm de}(z)$ crossing the phantom barrier. 
In fact, this
is the case with many parametrizations adjusted to fit SNIa 
data \cite{Lazkoz:2005sp}.

The existence of a phantom barrier is hinted in our approach by
the presence of the
term $\dot{w}_{\rm de}\left(1+w_{\rm de}\right)^{-1}$ in 
Eq.~(\ref{lin-de}). Although the divergence at $w_{\rm de}=-1$ is
not necessarily fatal for the solutions of the differential equations,
here we consider only dark energy parametrizations that
are phantom or non-phantom during all times.

We will study a parametrization of the dark energy EoS of the form
\cite{Chevallier:2000qy}:
\begin{equation}
w_{\rm de} = 
w_{0}+w_{1} (1-a) =
w_{0}+w_{1} \frac{z}{1+z} \; ,
\end{equation}
and we choose parameters $w_0$ and $w_1$ which are
consistent with the $2\sigma$ regions which are jointly
constrained by observations
of the CMB, supernovas and baryon oscillations
-- see, for instance, \cite{Wang:2007mz,Wright:2007vr}.

\begin{figure}[htb]
\includegraphics[scale=0.31]{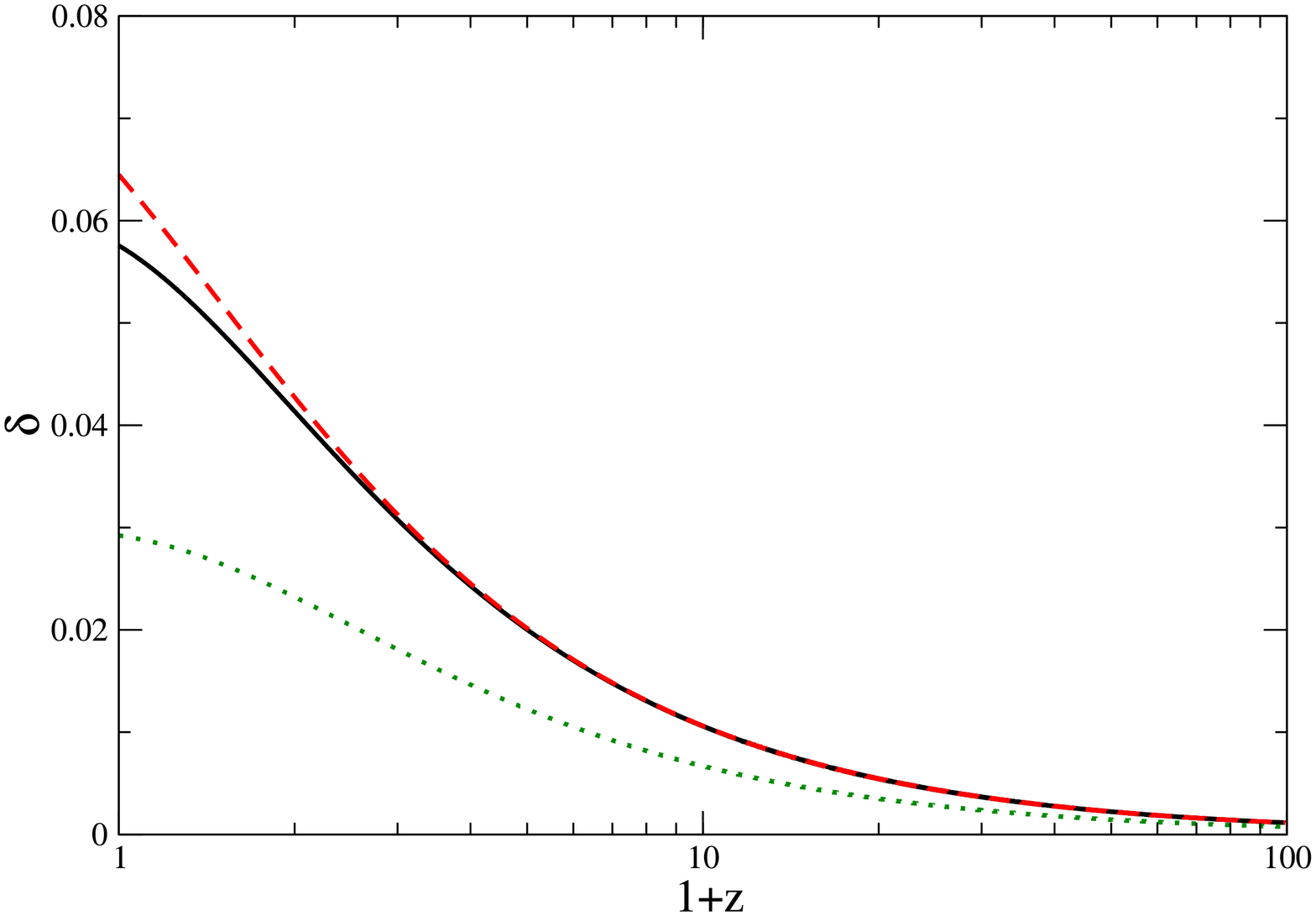}
\hspace{-1cm}
\includegraphics[scale=0.31]{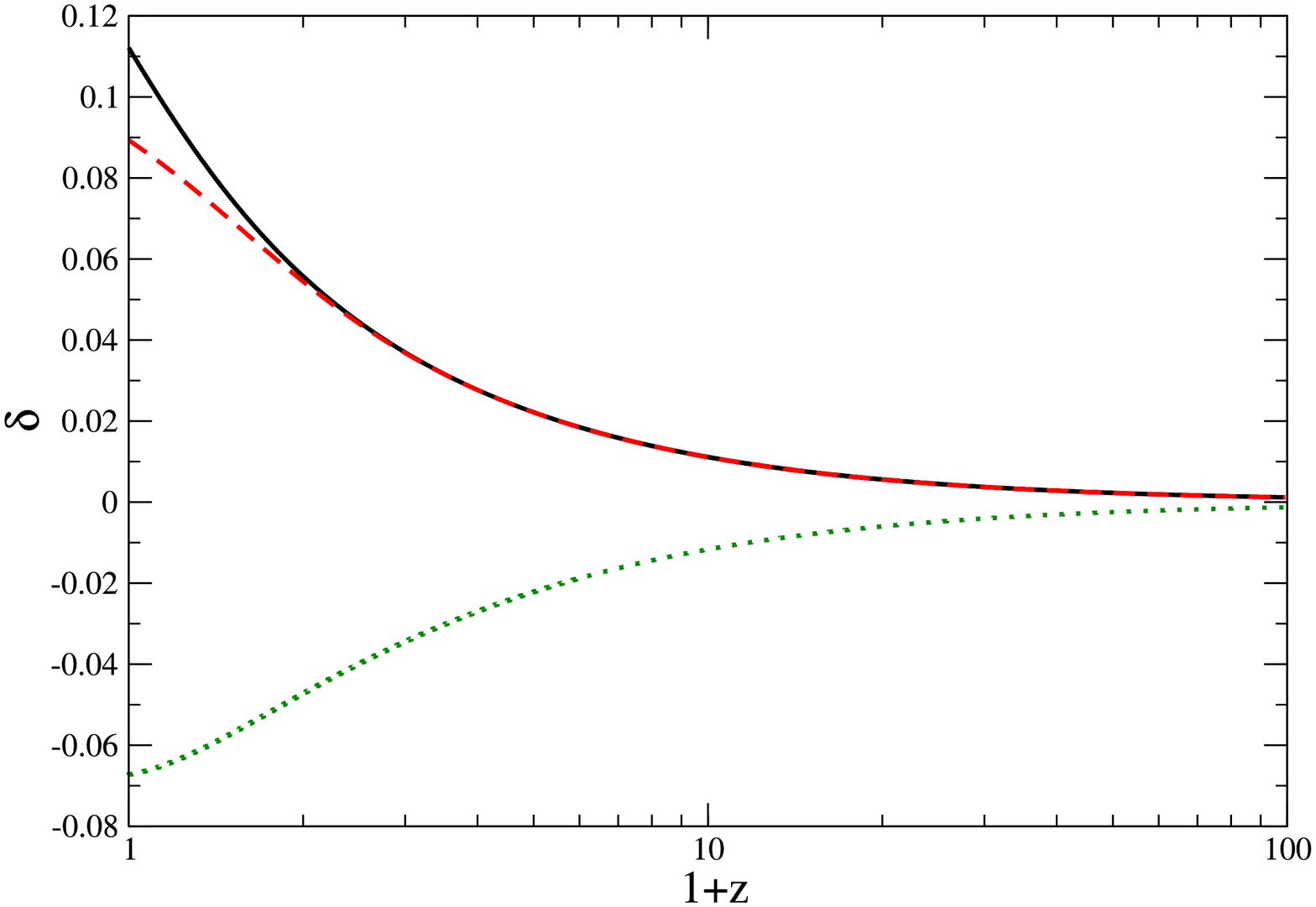}
\vspace{-0.8cm}
\caption{\sf Evolution of $\delta_{\rm m}\left(z\right)$ in linear regime with dark
energy adiabatic IC with $w_0 = -0.75$ and $w_1 = 0.4$ (non-phantom, left figure)
and with $w_0 = -1.1$ and $w_1 = -1$  (phantom, right figure)
including dark energy perturbations (full line) and without
dark energy perturbations (dashed line). The
growth of dark energy perturbations (dotted line) is also included.
\label{lin_variable}}
\end{figure}

In Fig.~\ref{lin_variable} we show the impact of 
dark energy flucutations
for a variable EoS for both phantom and non-phantom 
cases. The results are similar to the constant EoS cases.

\section{Non-linear regime}

In the non-linear regime, as in the linear regime, 
we consider again only models that are phantom or 
non-phantom at all times.
We solve Eqs.~(\ref{system1})-(\ref{system2}), and
for brevity's sake we limit our scope to the parametrizations
discussed in subsection \ref{variable}: $(w_0,w_1)=(-0.75,0.4)$
and $(-1.1,-1)$.
In particular, we are interested in the impact of the dark energy
fluctuations on the collapse of dark matter structures.
We will see that the effects found in the linear case (of the order of a 
few percent) are amplified by the non-linear evolution.

\begin{figure}[htb]
\includegraphics[scale=0.31]{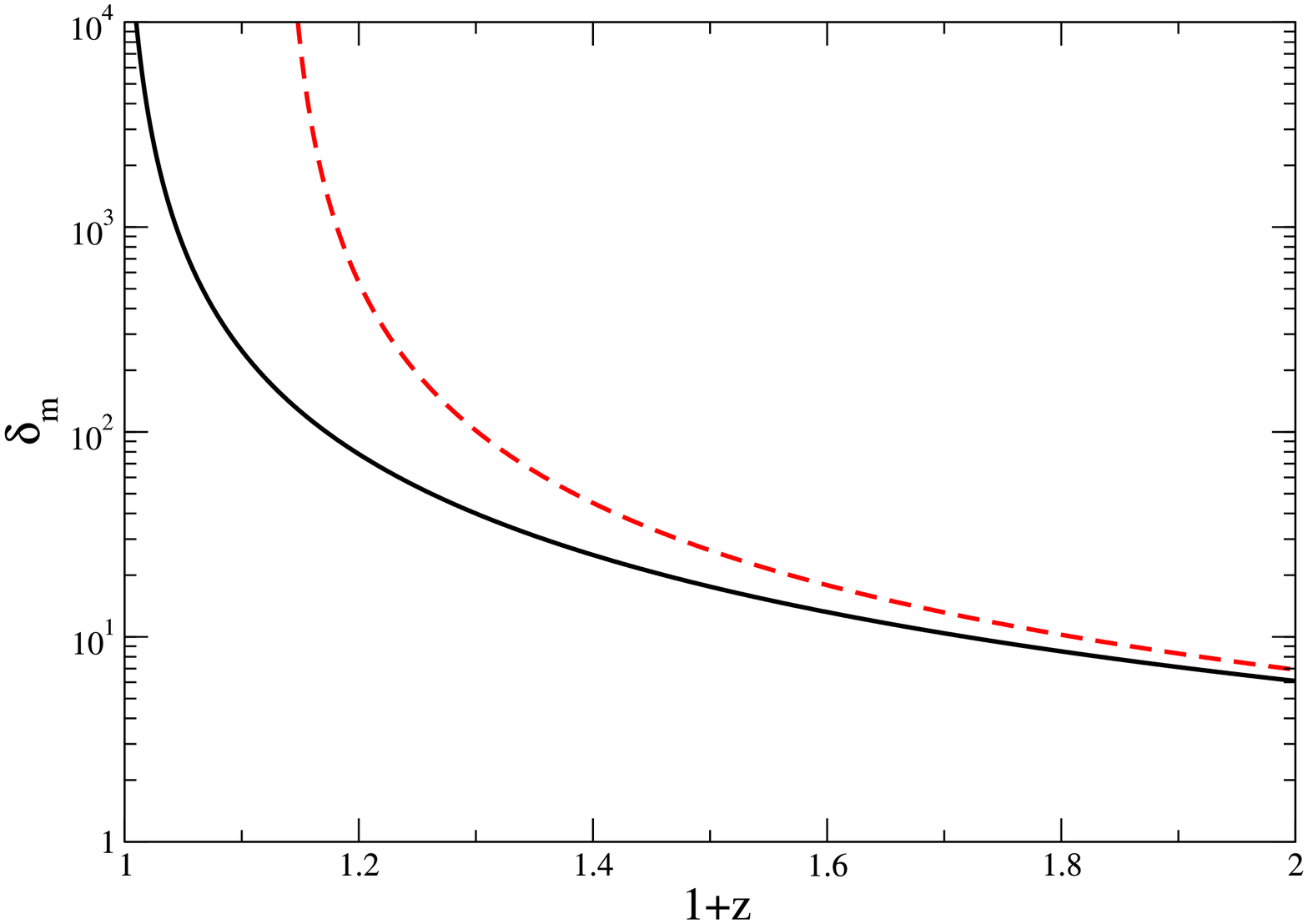}
\hspace{-1cm}
\includegraphics[scale=0.31]{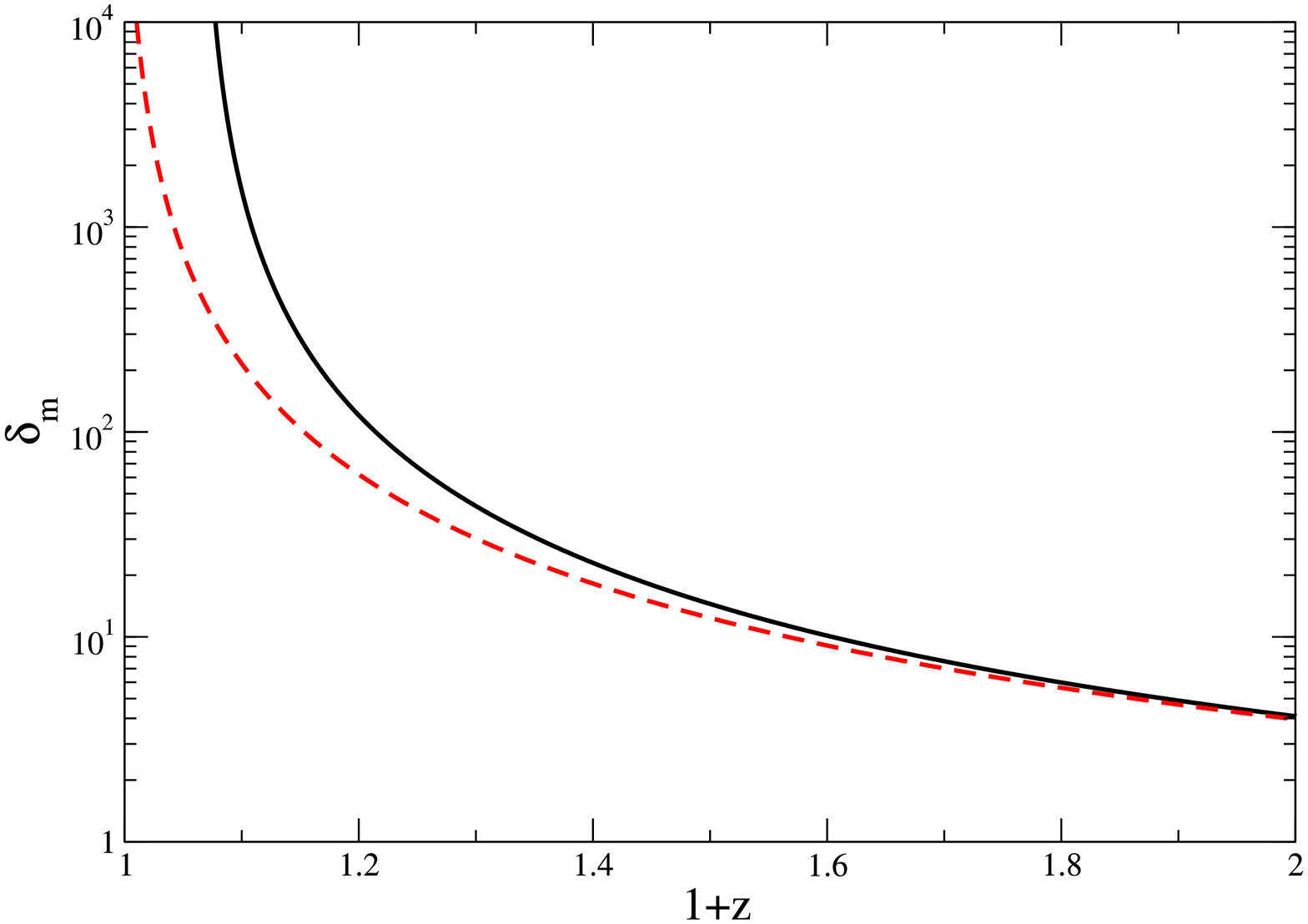}
\vspace{-1cm}
\caption{\sf Non-linear evolution of $\delta_m\left(z\right)$ for dark matter with (solid) and without (dashed) dark energy perturbation for the case of a non-phantom (left panel) and a phantom (right panel) 
parametrization. In the left panel, initial conditions
are chosen such that dark matter with dark energy perturbations
collapses today ($z=0$), whereas in the right panel initial conditions are chosen such that dark matter without dark energy perturbations collapses today. 
\label{collapse}}
\end{figure}

In Fig.~\ref{collapse} (left panel) we show the dark matter
density contrast for initial conditions chosen such that the
collapse (indicated by the divergence of the density contrast) of a 
spherical dark matter structure happens at the present time ($z=0$).
Physically, this would correspond to the formation of an object
such as a supercluster.
Dark energy fluctuations have a dramatic effect in this case: 
in the non-phantom case, that structure would 
have collapsed much earlier if the dark energy fluctuations
had not been taken into account.
In the right panel of Fig.~\ref{collapse} we also show a phantom 
parametrization, where 
initial conditions are chosen such that
the collapse of the dark matter structure takes place today
without dark energy perturbations. In this case, 
the inclusion of dark energy perturbations enhances
the clustering of dark matter and cause that same structure
to collapse earlier.

\begin{figure}[htb]
\includegraphics[scale=0.31]{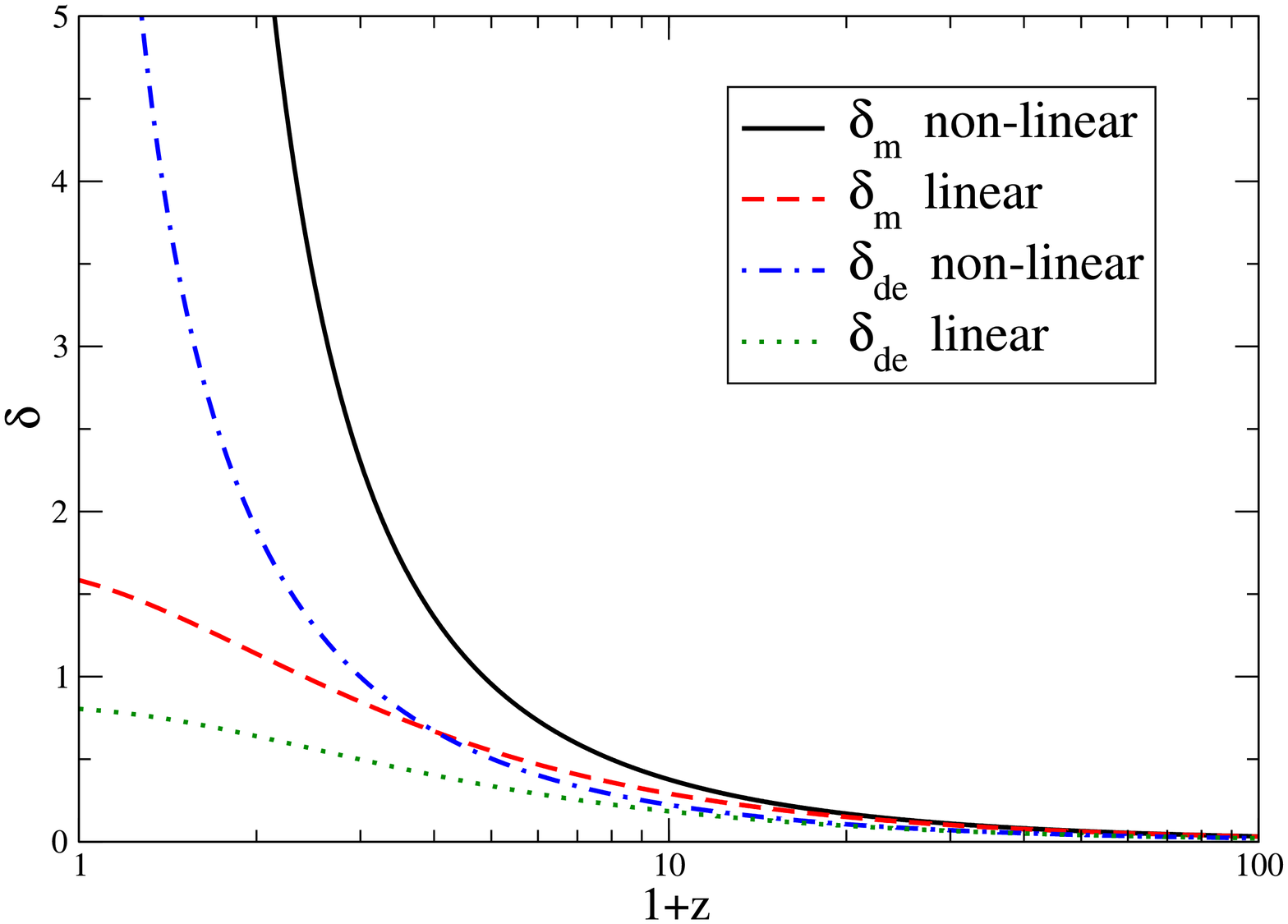}
\hspace{-1cm}
\includegraphics[scale=0.31]{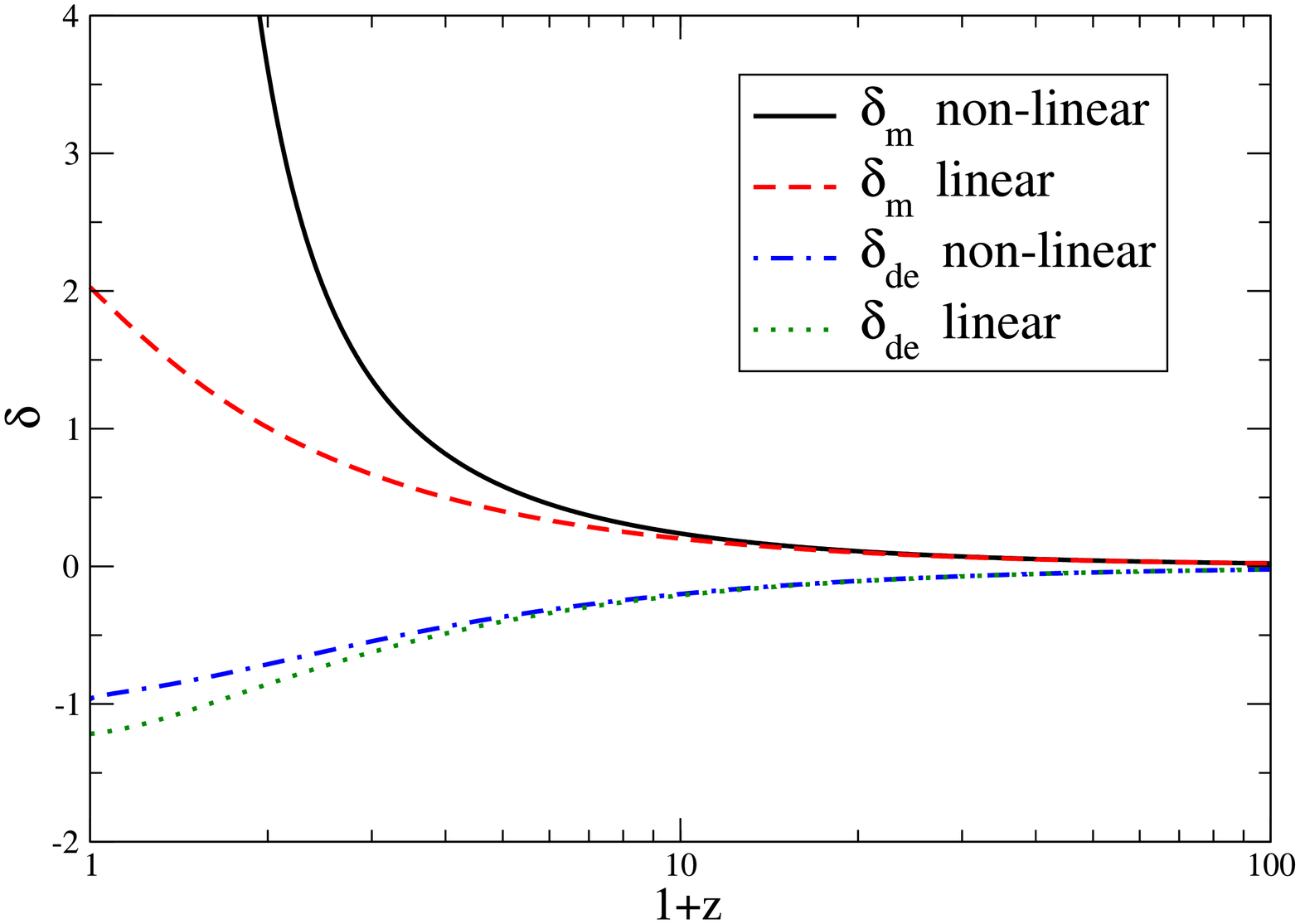}
\vspace{-1cm}
\caption{\sf Non-linear evolution of $\delta\left(z\right)$ for dark matter (solid line)
and dark energy (dot-dashed line). The initial conditions are chosen such that
dark matter collapses today ($z=0$). Also shown are the linearized evolution of
dark matter (dashed line) and dark energy (dotted line) perturbations.
The left and right panels correspond to the parametrizations 
$(w_0,w_1)=(-0.75,0.4)$ (left panel) and $(-1.1,-1)$ (right panel.)
\label{nonlinear}}
\end{figure}

In Fig.~\ref{nonlinear} we contrast the evolution of the non-linear
densities in dark matter and dark energy against the evolution
of the linearized densities. In that figure the initial
conditions are chosen such that the non-linear
dark matter perturbation diverges at $z=0$. Hence, the 
value of the linearized dark matter perturbation at $z=0$ in this 
case corresponds to the definition
of the critical density contrast for that redshift, $\delta_c(z=0)$. In 
the next Section we will employ the critical density contrast as a
function of redshift in order to
to compute the Press-Schechter function.

\section{Number Counts of Dark Matter Halos}\label{s.mf}
$ $

We will use the Press-Schechter approach \cite{Press:1973iz}
in order to estimate the number counts of dark 
matter halos for different bins of redshifts and halo masses. More realistic 
mass functions could be used (see, for instance, \cite{Sheth:1999su})
but our intention here is simply to point out how 
number counts differ in scenarios with dark energy
compared to the standard $\Lambda$CDM model. We 
believe these differences would be essentially 
the same in more sophisticated models 
of nonlinear structure formation, such as
the model of ellipsoidal collapse.

The Press-Schechter formalism assumes
that the fraction of mass in the universe contained in
gravitationally bound systems with masses greater than $M$ is
given by the fraction of space where the linearly evolved density
contrast exceeds a threshold $\delta_c$, and that the density contrast 
is normally distributed with zero mean and variance $\sigma^2(M)$ -- 
the root-mean-square value of the density contrast $\delta$ at scales 
containing a mass $M$. Therefore, it is assumed that for 
a massive sphere 
to undergo gravitational collapse at a redshift $z$, 
its {\it linear} overdensity
should exceed a certain threshold $\delta_c$, defined
as the linearly evolved density contrast at the instant
when the non-linear density contrast associated with the
mass $M$ diverges({\it i.e.}, at the moment of collapse.) 
Since each scale $M$ collapses at a given redshift (bigger
masses, corresponding to larger scales, collapse later), the
critical density contrast is a function of redshift, 
$\delta_c=\delta_c(z)$.
Nevertheless, notice that only linear quantities 
are used in this formalism. For a review of the 
Press-Schechter formalism see, for instance, 
\cite{Peacock}.

\begin{figure}[h]
   \centerline{\psfig{file=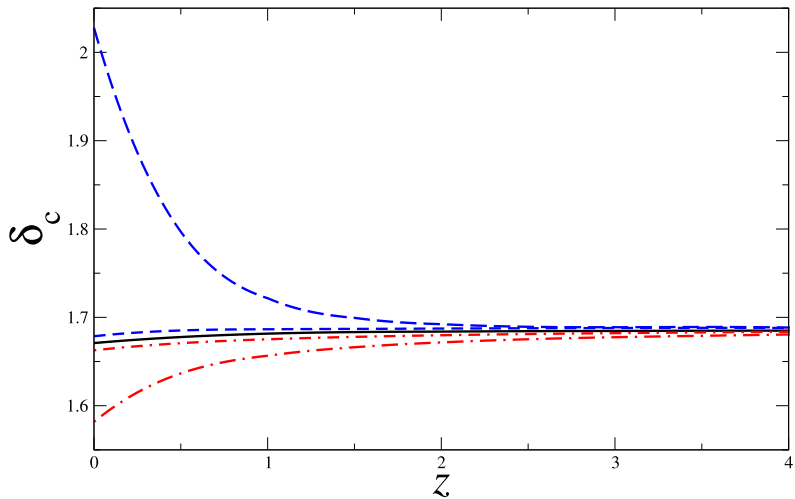,scale=0.4,angle=0}}
   \caption{\sf Values of the linear critical density 
contrast $\delta_c$
as a function of the collapse redshift.
The solid line is the usual $\Lambda$CDM result.  
The values which do not include dark energy perturbations are shown
for phantom (dashed line) and non-phantom 
(dot-dashed line) cases.
The values with dark energy perturbations are shown for 
phantom (long dashed line)
and non-phantom (long dot-dashed line).
The non-phantom and phantom cases correspond to the same 
parametrizations used
in Fig.~\ref{lin_variable}.}
\label{deltac}
\end{figure}

In order to illustrate how dark energy affects gravitational
collapse, in Fig.~\ref{deltac} we show $\delta_c(z)$ in a few
dark energy scenarios. Notice that the inclusion of
dark energy perturbations has a dramatic effect in $\delta_c(z)$, with a
substantial suppression in the non-phantom case and a substantial
enhancement in the phantom case. If dark energy perturbations are
not included we reproduce the results of \cite{Liberato:2006un}.

These assumptions lead to the well-known analytical 
formula for the comoving 
number density of collapsed halos of mass in the range 
$M$ and $M+ dM$ at a 
given redshift $z$: \be
   \frac{dn}{dM}(M,z) = -\sqrt{\frac{2}{\pi}}\frac{\rho_{\rm m0}}{M}
   \frac{\dl_c(z)}{\sigma (M,z)} \frac{d\ln\sigma (M,z)}{d M}
   \exp\left[-\frac{\dl_c^2(z)}{2\sigma^2 (M,z)}\right] \,,
   \label{mf}
\ee where $\rho_{\rm m0}$ is the present matter 
density of the universe 
and $\delta_c(z)$ is the linearly extrapolated density 
threshold above which 
structures collapse, {\it i.e.}, $\delta_c(z) = \dl_{\rm lin} (z = z_{\rm col})$.

The quantity:
\be
   \sigma (M,z) = D(z)\sigma_M
   \label{sigmamz}
\ee is the linear theory {\it rms} density fluctuation in spheres
of comoving radius $R$ containing the mass $M$, where 
$D(z)\equiv \delta_m(z)/\delta_m(z=0)$ is the linear growth function 
obtained from Eq.~(\ref{lin-mat}). The smoothing scale $R$ is often 
specified by the mass within the volume defined by the window 
function at the present 
time, see {\it e.g.} \cite{Peebles:1994xt}. In our analysis we use 
the fit given by \cite{Viana:1995yv}:
\be
   \sigma_M = \sigma_8 \left(\frac{M}{M_8}\right)^{-\gamma(M)/3} \,,
   \label{sigmam}
\ee where $M_8 = 6 \times 10^{14} \omo h^{-1} M_{\odot}$ is the
mass inside a sphere of radius $R_8 = 8 h^{-1} {\rm Mpc}$, and
$\sigma_8$ is the variance of the over-density field smoothed on a
scale of size $R_8$. The index $\gamma$ is a function of the mass
scale and the so-called shape parameter, $\Gamma = \omo h\; e^{-\Omega_b -
\Omega_b/\omo}$ ($\Omega_b = 0.05$ is the baryonic density 
parameter) \cite{Viana:1995yv} :
\be
   \gamma (M) = (0.3 \, \Gamma + 0.2) \left[ 2.92 + \frac{1}{3} \log
   \left(\frac{M}{M_8}\right) \right] \,.
   \label{gamma}
\ee

For a fixed $\sigma_8$ (power spectrum normalization) the predicted
number density of dark matter halos given by the above formula is
uniquely affected by the dark energy models through the ratio
$\dl_c(z)/D(z)$. In order to compare the different models, we will
normalize to mass function to the same value today, that is, we
will require: 
\be \sigma_{8,Mod} =
    \frac{\delta_{c,Mod}(z=0)}{\delta_{c,\Lambda}(z=0)} 
\sigma_{8,\Lambda}\,,
\ee where the label $Mod$ indicates a given model and we use
$\sigma_{8,\Lambda} = 0.76$ \cite{Spergel:2006hy}.

The effect of dark energy on the number of dark matter halos is
studied by computing two quantities. The first one is the all sky
number of halos per unit of redshift, in a given mass bin:
\be
   {\cal N}^{\rm bin}\equiv\frac{dN}{dz} =
   \int_{4\pi} d\Omega\int_{M_{\rm inf}}^{M_{\rm sup}}
   \frac{d n}{d M} ~\frac{d V}{d z d\Omega} ~dM \,,
   \label{nbin}
\ee
where the comoving volume element is given by:
\be
   \frac{dV(z)}{dz d\Omega} = r^2(z)/H(z),
\ee where $r(z) = \int_0^z H^{-1}(x) dx$ is the comoving distance.
Note that the comoving volume element depends only 
on the cosmological background 
and is identical for models with and without perturbations 
in dark energy. 
The diferences in the number counts when one includes
dark energy clumping are due to modifications in
$\dl_c(z)$ and $D(z)$. 

\begin{figure}[htb]
   \centerline{
   \psfig{file=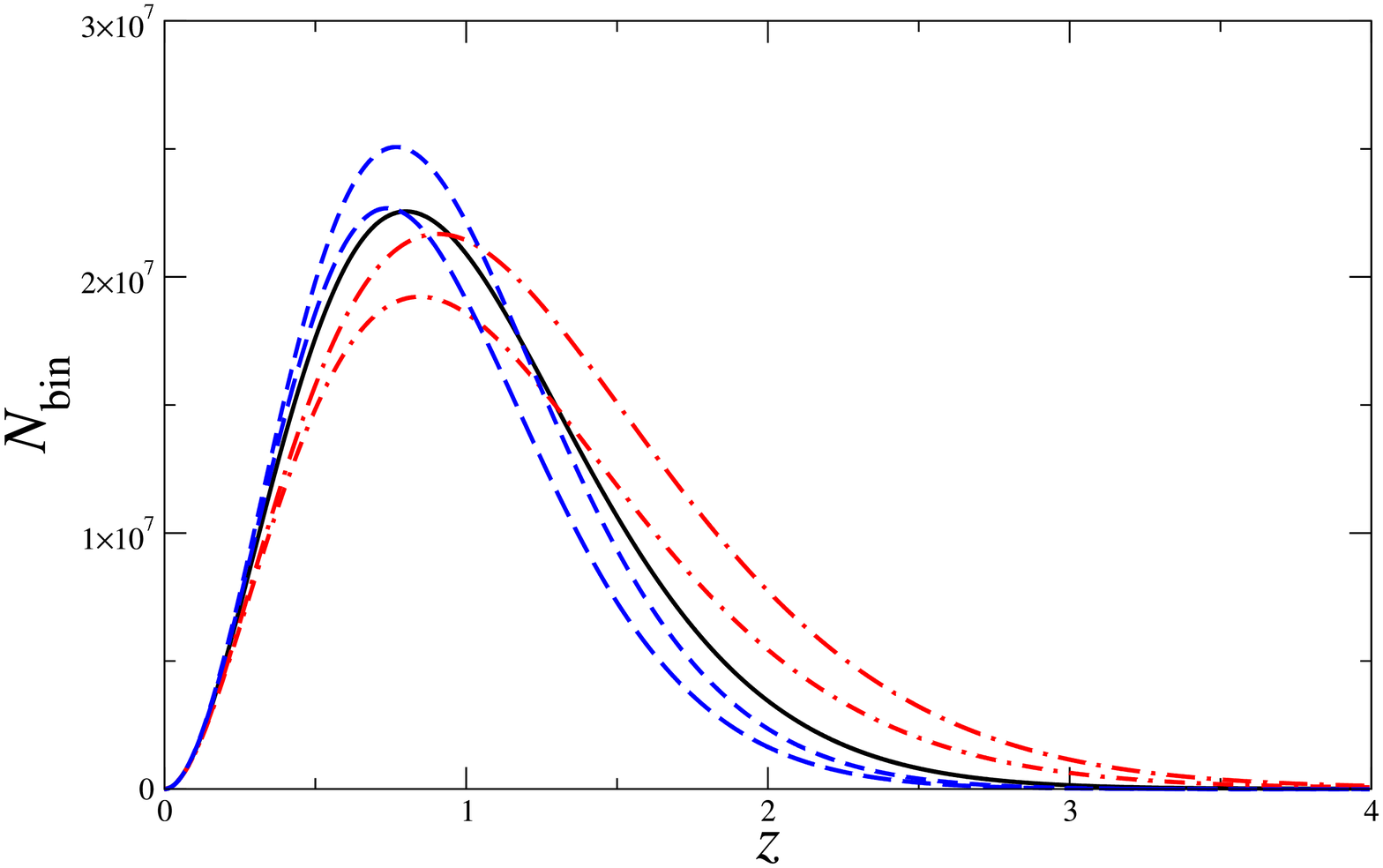,scale=0.3,angle=0}
   \psfig{file=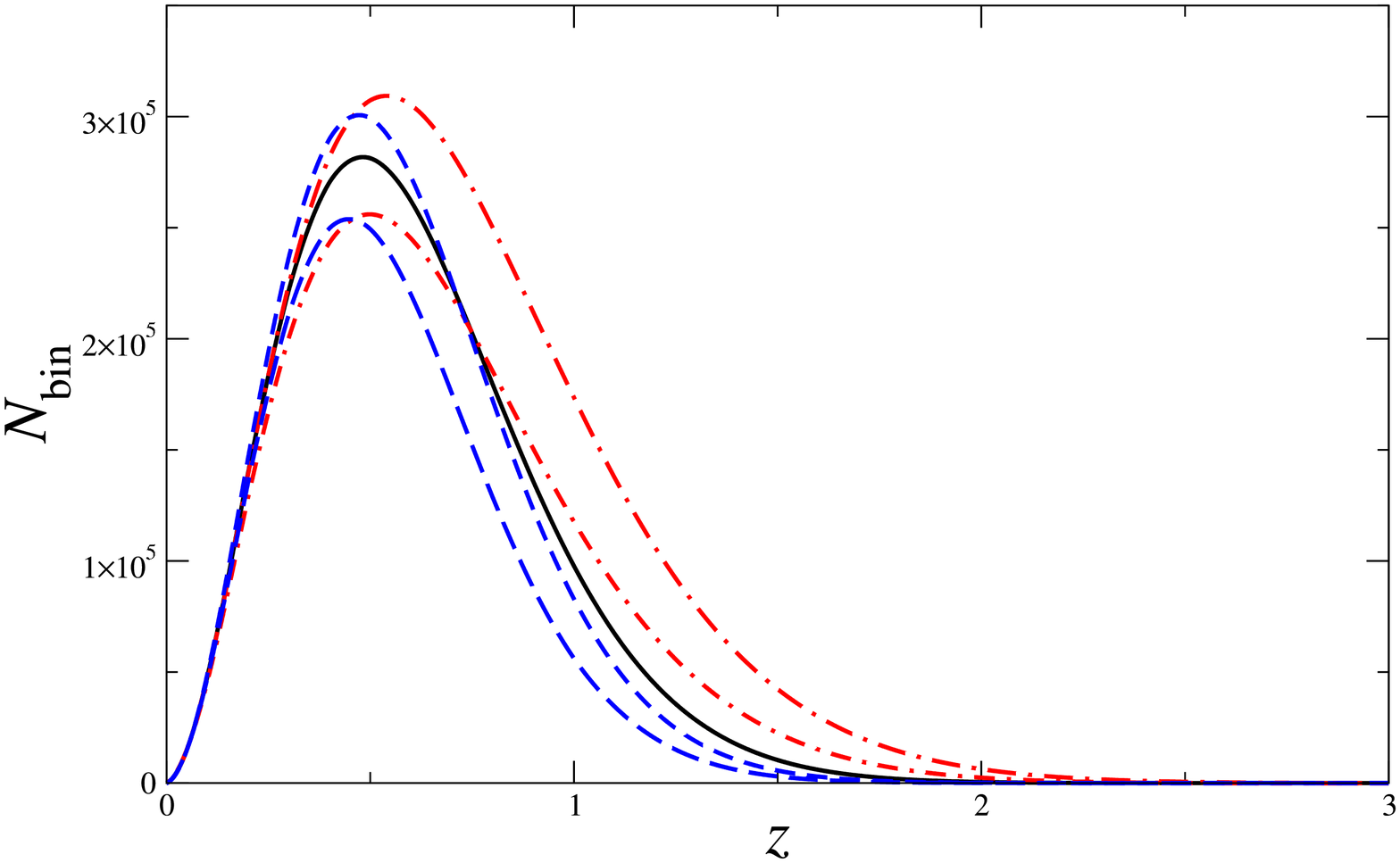,scale=0.3,angle=0}}
   \centerline{
   \psfig{file=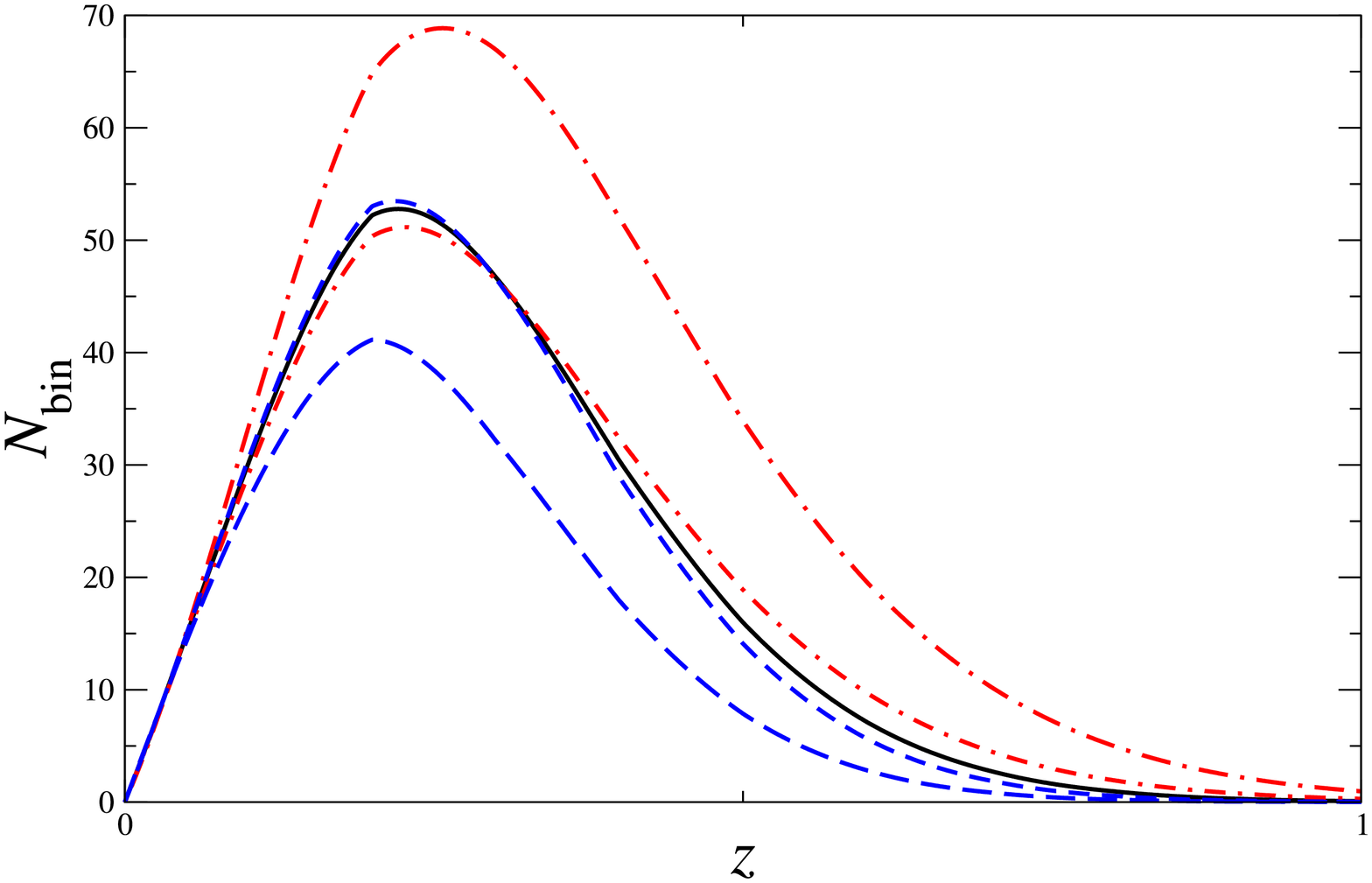,scale=0.3,angle=0}}
   \caption{\sf Evolution of number counts in mass bins with redshift
   for objects with masses within the range 
$10^{13}<M/(h^{-1}M_\odot )<10^{14}$ (top left panel), 
$10^{14}<M/(h^{-1}M_\odot )<10^{15}$ (top right panel) and
$10^{15}<M/(h^{-1}M_\odot )<10^{16}$ (bottom panel).
   The solid line corresponds to the fiducial $\Lambda$CDM result.  
   The number counts including dark energy perturbations are shown for 
   non-phantom (long dot-dashed line) and phantom (long dashed line) 
   models. 
   The results without the inclusion of dark energy perturbations are 
   also shown for non-phantom (dot-dashed line) and phantom 
   (dashed line) models.}
   \label{f.nbin}
\end{figure}


The second quantity we compute is the all sky integrated number
counts above a given mass threshold, $M_{\rm inf}$, and up to
redshift $z$ \cite{Nunes:2004wn}: \be
   N(z,M>M_{\rm inf})= \int_{4\pi}~d\Omega\int_{M_{\rm
   inf}}^{\infty}\int_0^z \frac{dn}{dM}~\frac{dV}{dz'd\Omega}~dM
   dz'\,.
   \label{nint}
\ee
Our knowledge of both these quantities for galaxy clusters will
improve enormously with upcoming cluster surveys operating at
different wavebands, such as the South Pole Telescope \cite{Ruhl:2004kv}.

We can now examine the modifications caused by a clustering
dark energy component on the number
of dark matter halos with the same observable computed in the
standard $\Lambda$CDM model. First we show how the 
different equations of state impact the number of dark matter halos in
given mass bins [$M_{\rm inf}, M_{\rm sup}$] typical of
the present-day cosmological structures, namely [$10^{13},10^{14}$]$h^{-1}M_\odot$,
[$10^{14},10^{15}$]$h^{-1}M_\odot$ and [$10^{15},10^{16}$] $h^{-1}M_\odot$.
The number counts in mass bins, ${\cal N}^{\rm
bin}=dN/dz$, obtained from (\ref{nbin}), are shown in Fig.
\ref{f.nbin}.
In each panel we plot the actual number
counts together with the number counts computed for a
fiducial $\Lambda$CDM model (solid lines), for each mass bin. 
Notice that the more massive
structures are less abundant and form at later times, as it should
be in the hierarchical model of structure formation. There is
a slight shift of the peak redshift for structure formation
in the distinct dark energy models considered.
The differences with respect to the $\Lambda$CDM model become more
significant in the bins with larger masses -- but, of course, given the
small number of such massive objects, the uncertainty due to shot 
noise also becomes increasingly important.

\begin{figure}[h]
   \centerline{
   \psfig{file=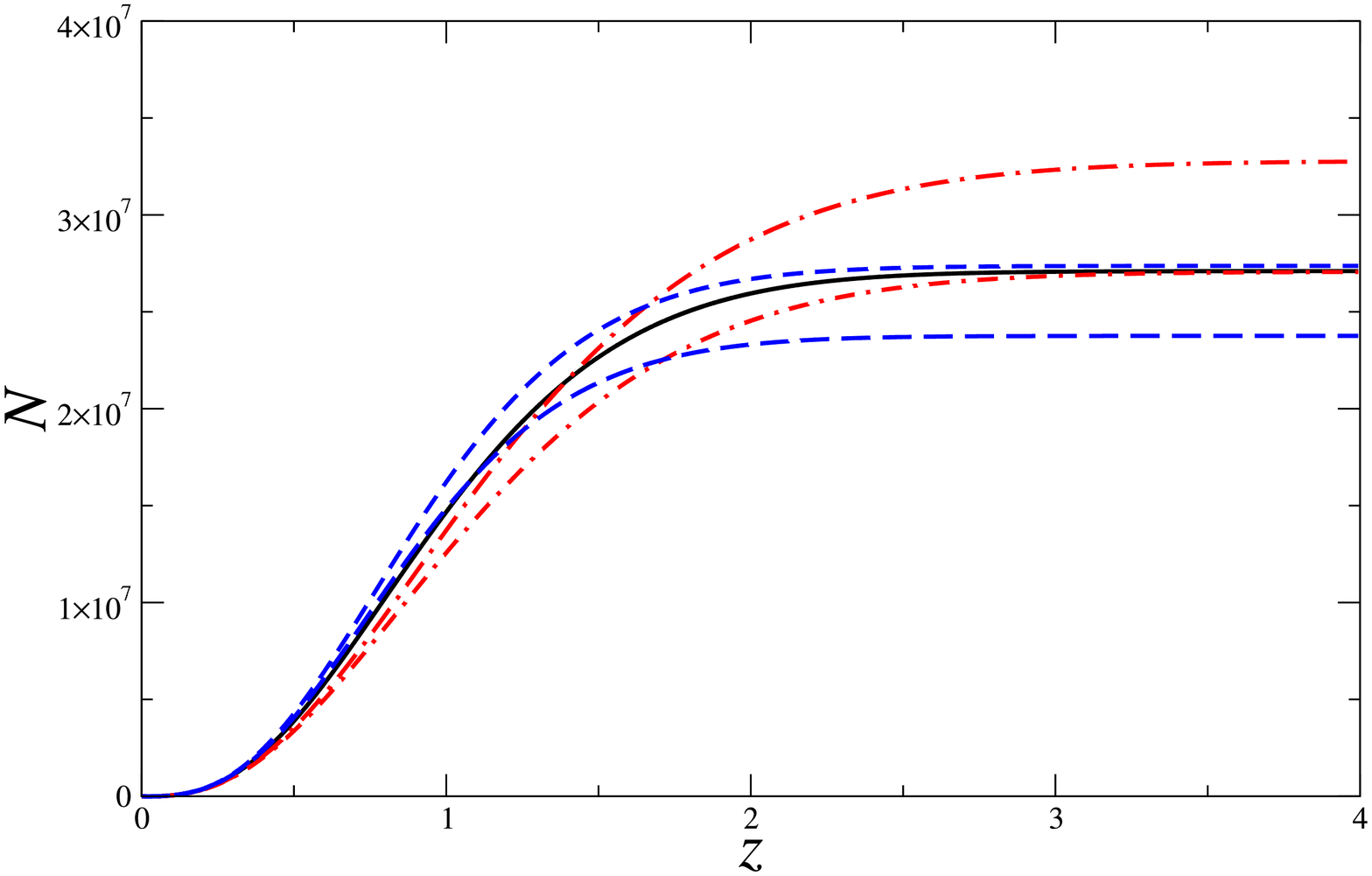,scale=0.3,angle=0}
   \psfig{file=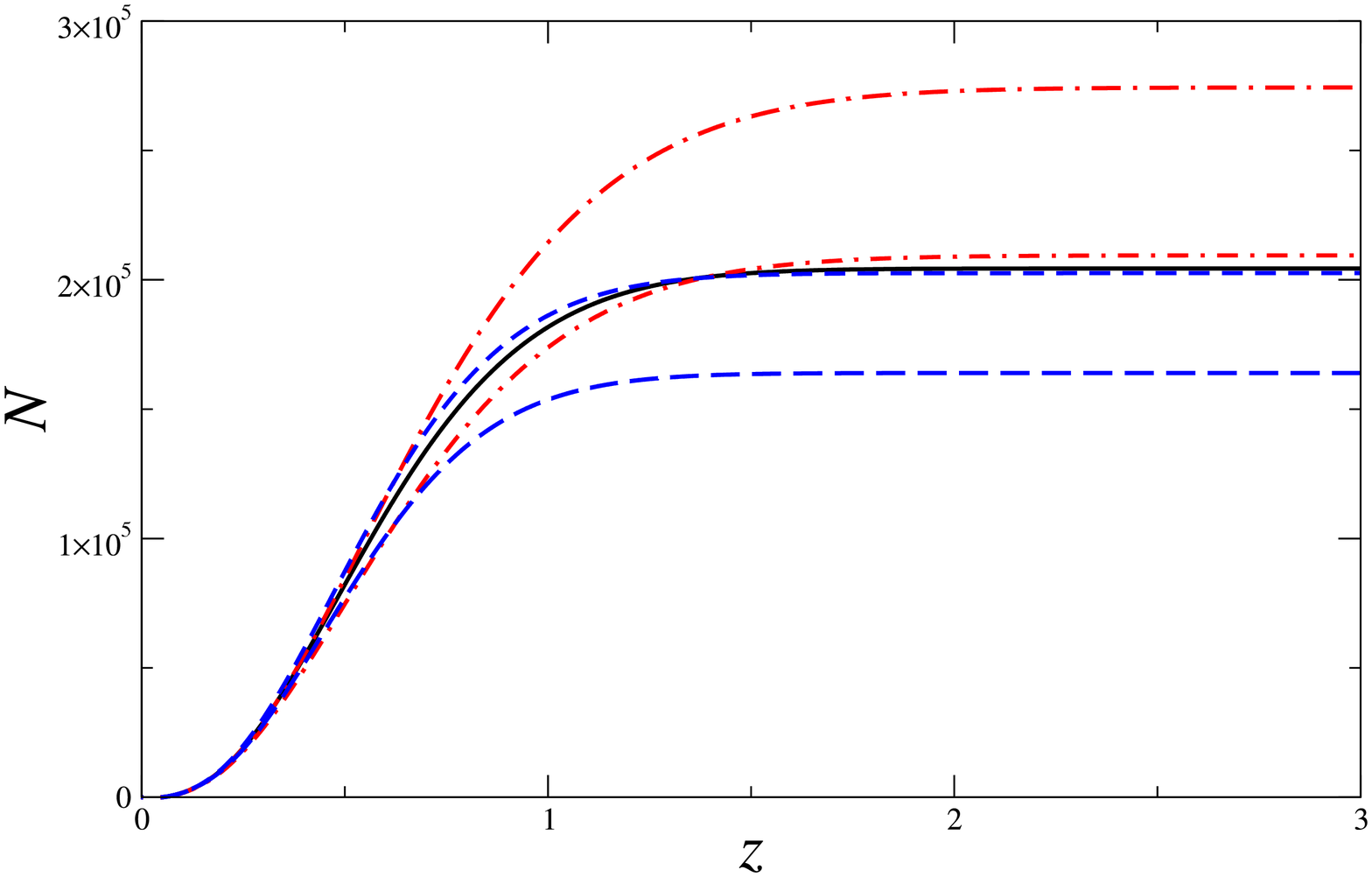,scale=0.3,angle=0}}
   \centerline{\psfig{file=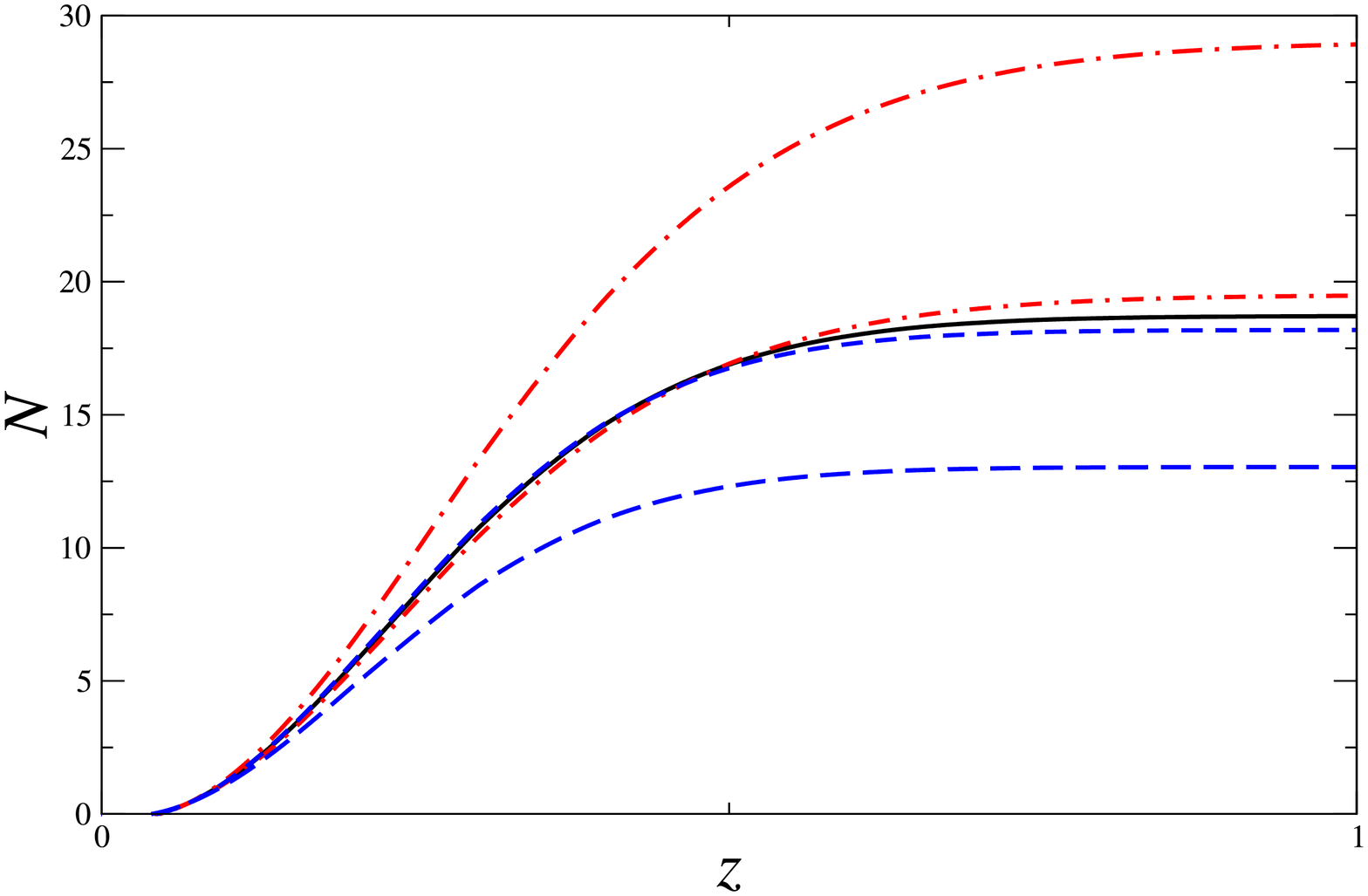,scale=0.3,angle=0}}
   \caption{\sf Evolution of the integrated number counts up to redshift $z$ 
   for objects with  
   $M> 10^{13} h^{-1}M_\odot$ (top left panel),
   $M> 10^{14} h^{-1}M_\odot$ (top right panel) and
   $M> 10^{15} h^{-1}M_\odot$ (bottom panel). The lines 
   correspond to the same cases as in Fig.~\ref{f.nbin}.}
   \label{f.nc}
\end{figure}

Another important observable quantity is the integrated number of
collapsed structures above a given mass, Eq.~(\ref{nint}). We
present results for the integrated number counts of structures
with masses above $10^{13}h^{-1}M_\odot$, $10^{14}h^{-1}M_\odot$,
and $10^{15}h^{-1}M_\odot$ (we 
cut-off the integration at $M_{\rm sup} = 10^{18} h^{-1}M_\odot$,
as such gigantic structures could not in practice be resolved today.)
 The results are displayed in Fig.~\ref{f.nc}, 
always compared with the results for the fiducial $\Lambda$CDM
model (solid lines.)  Notice that the integrated
number has a plateau that reflects the epoch of structure
formation for a given mass. In other words, there is no formation
of structures with mass above $10^{13}h^{-1}M_\odot$,
$10^{14}h^{-1}M_\odot$, and $10^{15}h^{-1}M_\odot$ for redshifts
roughly above $z=2$, $1.5$ and $0.7$, respectively. Again we find
large differences compared to the $\Lambda$CDM model 
when dark energy perturbations are included.

\section{Conclusions}

Our main goal in this paper was to study the effects of
including dark energy perturbations in the evolution
of matter perturbations in the linear and in the non-linear regimes.
Since we do not know what dark energy really is, we developed
a formalism whereby we can directly use a parametrization of its equation
of state in order to address this issue.

We have shown that the spherical collapse and the pseudo-newtonian
approaches to the study the non-linear evolution of dark matter and dark
energy perturbations are equivalent when one adopts an effective speed of 
sound  $c_{{\rm eff}}^{2}=w$. In the languange of spherical collapse, this
is equivalent to assuming that the equation of state is the same inside 
and outside the collapsed region.

We found distinct behaviours in the evolution of the dark matter
perturbations for phantom and non-phantom forms of dark energy.
Inclusion of dark energy perturbations inhibits the growth of dark 
matter perturbations for the non-phantom case but it enhances this
growth in the phantom case. The reason is that 
dark matter overdensities lead to dark energy 
overdensities in the non-phantom case, but they lead to 
underdensities in the phantom case.
Due to its gravitationally repulsive nature, dark energy overdensities inhibit, while
dark energy underdensities help, the growth of dark matter perturbations.

This effect is small in the linear regime but becomes dramatic
when studying the collapsed regions that have formed more recently. 
In particular, we found
a large modification in the critical density $\delta_c(z)$ even for
moderatly low redshifts.

We used the Press-Schechter formalism to estimate the modifications
due to dark energy perturbations
in observational quantities such as number counts of galaxy clusters,
which reflect the formation and distribution of dark matter halos. 
We found that there are large deviations compared to the standard
$\Lambda$CDM model, which are more significant for the larger structures.
We expect that these large deviations are a general consequence of
taking into account the non-linear dark energy perturbations that are
gravitationally coupled to the dark matter perturbations. 

Although our use of the EoS to describe dark energy perturbations
clearly constitutes a particular case, our choice was guided by the 
equivalence between the
SC and PN approaches. However, we believe that more general models can still
be described consistently in both approaches \cite{us2007}. 
Hopefully future data on number counts of galaxy clusters will be able
to discriminate among different models of dark energy.

\section*{Acknowledgments}
We would like thank C. Quercellini, T. Padmanabhan and I. Waga for useful comments.
LL is supported by a CAPES doctoral fellowship; RCB thanks
FAPESP for a doctoral fellowship.
The works of LRA and RR are partially supported by 
research grants and fellowships from CNPq, and 
by a research grant from FAPESP (Proc. 04/13668-0.)

\appendix
\section{}

In this paper we have analyzed the non-linear evolution of dark energy
and dark matter perturbations using the PN. One can ask about the
validity of such a model when pressure is taken into account. Here we
explicitly show that, up to linear order, pseudo-newtonian equations
are in good agreement with general relativity perturbation theory.

The relativistic equation for the gauge invariant perturbation, for a single
perfect fluid can be written as \cite{Kodama:1985bj}:
\begin{equation}
\ddot{\delta}+2H\left[1-3\left(w-\frac{c_{s}^{2}}{2}\right)\right]\dot{\delta}+\frac{3H^{2}}{2}\left(3w^{2}-8w-1+6c_{s}^{2}\right)\delta=-\frac{k^{2}}{a^{2}}c_{{\rm eff}}^{2}\delta\;.
\end{equation}
Neglecting the term on RHS, which can be understood as a large scale
approximation or a top-hat profile, and taking $w=const=c_s^2$ we have:
\begin{equation}
\ddot{\delta}+2H\left[1-\frac{3w}{2}\right]\dot{\delta}+\frac{3H^{2}}{2}\left(3w^{2}-2w-1\right)\delta=0\;,
\end{equation}
using $H=2/3\left(1+w\right)t$:
\begin{equation}
\ddot{\delta}+\frac{4-6w}{3\left(1+w\right)}\frac{\dot{\delta}}{t}+\frac{6w^{2}-4w-2}{3\left(1+w\right)^{2}}\frac{\delta}{t^{2}}=0\;.
\end{equation}
The solution to the relativistic equation above can be written as:
\begin{equation}
\delta=C_{+}t^{2\left(1+3w\right)/3\left(1+w\right)}+C_{-}t^{-1+2w/1+w}\;.
\end{equation}

In PN, Eq.~(\ref{newt_sc}), up to linear terms, with $c_{{\rm eff}}^{2}=w=const$
and $\delta_{PN}$ with no spatial dependence, is written as:
\begin{equation}
\ddot{\delta}_{PN}+2H\dot{\delta}_{PN}-\frac{3H^{2}}{2}\left(1+w\right)\left(1+3w\right)\delta_{PN}=0\;.
\end{equation}
and its solution is:
\begin{equation}
\delta_{PN}=C_{+}t^{2\left(1+3w\right)/3\left(1+w\right)}+C_{-}t^{-1}\;.
\end{equation}

The relativistic and pseudo-newtonian solutions differ only in the
decaying mode. Note that for dust, $w=0$, the two solutions coincide,
even for the SC model \cite{Gaztanaga:2000vw,Fosalba:1997tn,Hwang:2005xt}.
Although both theories seem to be in good agreement, we must be
watchful of phantom models. In this case the usual relativistic decaying mode
$\sim t^{-1+2w/1+w}$ becomes a growing mode, while in PN it is always
decaying. This behaviour is expected when phantom dark energy starts
to dominate and shows that growing perturbations phantom dark energy
increase the gravitational potencial, unlike one should expect from
a homogeneous phantom dark energy model. For all other values of $-1<w\leq1$
the relativistic solution $\sim t^{-1+2w/1+w}$ is a decaying mode.
Hence, if the sub-dominant mode is not irrelevant for some reason,
then the PN may not be reliable for phantom models at times when dark energy
is strongly dominating the background evolution, i.e, $\Omega_{\rm de}\approx1$.
Fortunately, this situation does not arise in our study.

\section*{References}
\bibliographystyle{h-physrev3}
\bibliography{paper_v5}

\end{document}